\documentclass[structabstract]{aa}

\usepackage{graphicx}
\usepackage{longtable}
\usepackage[varg]{txfonts}

\begin{document}

   \title{The mid-infrared diameter of W~Hydrae\thanks{Based on observations made with the Very Large Telescope Interferometer (VLTI) at the Paranal Observatory under program IDs 079.D-0140, 080.D-0005, 081.D-0198, 082.D-0641 and 083.D-0294.}$^{,}$\thanks{Color versions of the figures are available in electronic form via http://www.aanda.org}}
   
   %\footnote{Color versions of the figures are available in electronic form via http://www.aanda.org}

   %\subtitle{...}
 
   \author{R. Zhao-Geisler\inst{1}\fnmsep\thanks{Fellow of the International Max Planck Research School (IMPRS),
              \email{rgeisler@lsw.uni-heidelberg.de}}
          \and A. Quirrenbach\inst{1}
          \and R. K\"ohler\inst{1,3}
          \and B. Lopez\inst{2}
          \and C. Leinert\inst{3}
          }

   \institute{Zentrum f\"ur Astronomie der Universit\"at Heidelberg (ZAH),
              Landessternwarte, K\"onigstuhl 12, D-69120 Heidelberg
    \and
              Laboratoire Fizeau, Observatoire de la C$\hat{\mathrm{o}}$te d'Azur, BP. 4229, 06304 Nice Cedex 4, France
    \and
              Max-Planck-Institut f\"ur Astronomie, K\"onigstuhl 17, D-69120 Heidelberg
              }

   \date{Received 14 December 2010 / Accepted 17 March 2011}

%###########################################################################################
%###########################################################################################

  \abstract % 5 {} token are mandatory
  % context heading (optional)
   {}
  % aims heading (mandatory)
   {Asymptotic giant branch (AGB) stars are among the largest distributors of dust into the interstellar medium, and it is therefore important to understand the dust formation process and sequence in their strongly pulsating extended atmosphere. By monitoring the AGB star W~Hya interferometrically over a few pulsations cycles, the upper atmospheric layers can be studied to obtain information on their chemical gas and dust composition and their intracycle and cycle-to-cycle behavior.}
   %This leads to a better understanding of the complex dynamical atmospheric processes attributed to AGB stars.
  % methods heading (mandatory)
   {Mid-infrared ($8-13$~$\mu$m) interferometric data of W~Hya were obtained with MIDI/VLTI between April~2007 and September~2009, covering nearly three pulsation cycles. The spectrally dispersed visibility data of all 75 observations were analyzed by fitting a circular fully limb-darkened disk (FDD) model to all data and individual pulsation phases. Asymmetries were studied with an elliptical FDD.}
  % results heading (mandatory)
   {Modeling results in an apparent angular FDD diameter of W~Hya of about (80~$\pm$~1.2)~mas (7.8~AU) between 8 and 10~$\mu$m, which corresponds to an about 1.9~times larger diameter than the photospheric one. The diameter gradually increases up to (105~$\pm$~1.2)~mas (10.3~AU) at 12~$\mu$m. In contrast, the FDD relative flux fraction decreases from (0.85~$\pm$~0.02) to (0.77~$\pm$~0.02), reflecting the increased flux contribution from a fully resolved surrounding silicate dust shell. The asymmetric character of the extended structure could be confirmed. An elliptical FDD yields a position angle of (11~$\pm$~20)$^\circ$ and an axis ratio of (0.87~$\pm$~0.07). A weak pulsation dependency is revealed with a diameter increase of (5.4~$\pm$~1.8)~mas between visual minimum and maximum, while detected cycle-to-cycle variations are smaller.}
  % conclusions heading (optional)
   {W~Hya's diameter shows a behavior that is very similar to the Mira stars RR~Sco and S~Ori and can be described by an analogous model. The constant diameter part results from a partially resolved stellar disk, including a close molecular layer of H$_2$O, while the increase beyond 10~$\mu$m can most likely be attributed to the contribution of a spatially resolved nearby Al$_2$O$_3$ dust shell. Probably due to the low mass-loss rate, close Fe-free silicate dust could not be detected. The results suggest that the formation of amorphous Al$_2$O$_3$ occurs mainly at visual minimum. A possible close Al$_2$O$_3$ dust shell has now been revealed in a few objects calling for self-consistent dynamic atmospheric models including dust formation close to the star. The asymmetry might be explained by an enhanced dust concentration along an N-S axis.}

   \keywords{stars: AGB and post-AGB --
             stars: individual: W~Hya --
             stars: circumstellar matter --
             stars: diameter --
             infrared: stars --
             techniques: interferometric}

   \titlerunning{The mid-infrared diameter of W~Hydrae}
             
   \maketitle

%###########################################################################################
%###########################################################################################

\section{Introduction}\label{secIntro}

    Asymptotic giant branch (AGB) stars have high luminosities, a high mass-loss rate, and a relatively cool atmospheric temperature, and they represent a late stage of stellar evolution. Most of the AGB stars pulsate (as Miras, semi-regular variables or irregular variables) and have typical main-sequence progenitor masses of 0.8 to 10~M$_\odot$. A~large fraction of the mass is concentrated in a tiny degenerate carbon-oxygen core, surrounded by a helium- and hydrogen-burning shell and an extended stellar envelope with a very extended convective zone. Photospheric diameters are typically a few AU. The star is embedded in a dusty circumstellar envelope (CSE) (cf.~e.g.~review book by Habing \& Olofsson \cite{Habing_Olofsson_04}).

    The M-type (O-rich) star \object{W~Hya} is interferometrically observed in the thermal N-band to resolve the upper atmospheric layers over a few pulsations cycles. This gives the possibility of obtaining information on its chemical gas and dust composition and its behavior throughout and between pulsation cycles. This can lead to a better understanding of the complex dynamical atmospheric processes attributed to AGB stars. Since AGB stars are one of the most important distributors of dust into the interstellar medium, besides red super giant (RSG) stars and supernovae (SN), it is especially important to understand the dust formation process and sequence in their strongly pulsating extended atmosphere.
    
    In addition, asymmetries (e.g.~oblateness due to rotation or a companion, or non-symmetric brightness distributions due to stellar spots or large convection zones), which are not uncommon for AGB stars, can be investigated. Past asymmetry determinations of W~Hya were not very conclusive, in the sense that position angles over a wide range were reported, ranging from about $70^\circ$ (Szymczak et al.~\cite{Szymczak_et98}) in the radio to about $143^\circ$ (Lattanzi et al.~\cite{Lattanzi97}) in the visual, while no departures from symmetry could be detected in the near-infrared (IR) within the measurement uncertainties (Ireland et al.~\cite{Ireland04} and Monnier et al.~\cite{Monnier04}).

    W~Hya is one of the best observed AGB stars in the southern hemisphere. In particular, many interferometric diameter measurements at visual and near-IR wavelengths have been carried out (e.g.~Ireland et al.~\cite{Ireland04} and Woodruff et al.~\cite{Woodruff09}, cf.~Sect.~\ref{secPhaseSubWave}). Because of the very extended atmosphere, it is expected that different wavelengths probe different atmospheric layers in AGB stars (Baschek et al.~\cite{Baschek_et91} and Scholz et al.~\cite{Scholz01}). The visual and near-IR diameters of the star W~Hya range from about 30 up to 70~mas, correlated to the absorption and emission bands of the most abundant and radiatively important molecular species, such as H$_2$O, OH, CO, TiO, SiO, CO$_2$ and SO$_2$, besides H$_2$ (Hofmann et al.~\cite{Hofmann_et98} and Jacob et al.~\cite{Jacob_et00}). Therefore it is difficult to measure a continuum diameter, while these opaque molecular layers can easily have radii twice the continuum radius (e.g.~Mennesson et al.~\cite{Mennesson_et02}, Tej et al.~\cite{Tej_et03} and Ohnaka \cite{Ohnaka04}) with temperatures of 1000 to 2000~K, i.e.~below the temperature of the continuum-forming surface of the star, but higher than the temperature of the surrounding circumstellar dust envelope.

    Dust grains with high sublimation temperatures are already present in the outer atmosphere and are the seed particles for further dust growth (e.g.~Lorenz-Martins \& Pompeia \cite{LorenzMartins_Pompeia00} and Verhoelst et al.~\cite{Verhoelst_et09} and references therein). Radiative pressure leads then at larger radii to an acceleration of the dust and the gas (through frictional coupling), resulting in the characteristic high mass-loss rates of AGB stars (e.g.~Woitke \cite{Woitke06} and H\"ofner \cite{Hoefner08}). Several attempts have been made to include different dust species, such as aluminum oxide (Al$_2$O$_3$), silicates (e.g.~Mg$_2$SiO$_4$), spinel (MgAl$_2$O$_4$), and olivine (MgFeSiO$_4$), to model the spectral energy distribution and to explain the nearly two times larger mid-IR diameters by adding a dust shell to a dust-free atmospheric model (cf.~e.g.~for Mira stars Ohnaka et al.~\cite{Ohnaka_et05} and Wittkowski et al.~\cite{Wittkowski_et07}, and for RSG stars Perrin et al.~\cite{Perrin_et07} and Verhoelst et al.~\cite{Verhoelst_et09}). The radiative importance of dust in an extended molecular shell depends on the local temperature, the time dependent growth rate, and the wavelength-dependent optical properties. The former can be very different for different pulsation phases and cycles.
     
    Dynamical atmospheric and wind models for AGB stars were developed by Hofmann et al.~(\cite{Hofmann_et98}), Woitke et al.~(\cite{Woitke_et99}, \cite{Woitke06}), Scholz et al.~(\cite{Scholz01}), H\"ofner et al.~(\cite{Hoefner_et03}), Ireland \& Scholz \cite{Ireland_Scholz06}, H\"ofner \& Andersen (\cite{HoefnerAndersen_07}), Ireland et al.~(\cite{Ireland_et08}), Nowotny et al.~(\cite{Nowotny_et10}) and Lebzelter et al.~(\cite{Lebzelter_et10}). The exact chemical composition, pulsation phase, and pulsation cycle behavior of AGB stars, e.g.~the location of molecular and dust layers, is still very uncertain. Observations presented in this work can help to improve this situation and provide the opportunity to compare with theoretical model predictions.

    W~Hya is a large-amplitude, semi-regular variable (SRa, but sometimes also classified as Mira) with a pulsation period of about one year, and it is located in the P-L-diagram on sequence~C (fundamental mode pulsator, Lebzelter et al.~\cite{Lebzelter_et05}; sequence~1 in Riebel et al.~\cite{Riebel_et10}). The visual magnitude varies strongly, while the amplitude in the N-band is rather small (cf.~Sect.~\ref{secObsSubLC}). The radial velocity amplitude, derived from CO $\Delta\nu~=~3$ lines, is about 15 km$\,$s$^{-1}$ (Hinkle et al.~\cite{Hinkle_et97}, Lebzelter et al.~\cite{Lebzelter_et05}). Distance estimates to W~Hya range from 78~pc (Knapp et al.~\cite{Knapp_et03}, revised \emph{Hipparcos} value) to 115~pc (Perryman et al.~\cite{Perryman_et97}, \emph{Hipparcos} value). Throughout this paper an intermediate value of 98$^{+30}_{-18}$~pc from Vlemmings et al.~(\cite{Vlemmings_et03}) will be assumed (Very Long Baseline Interferometry maser measurement). W~Hya's luminosity is about 5400~$L_{\odot}$ (Justtanont et al.~\cite{Justtanont_et05}).

    After a description of the observational method and data reduction in Sect.~\ref{secObs}, the model is explained in Sect.~\ref{secMod}, where the results will be interpreted by the presence of dust close to a molecular layer. Sect.~\ref{secPhase} investigates the pulsation dependence of the apparent diameter, as well as departures from symmetry. A summary is given in Sect.~\ref{secConc}.

%###########################################################################################
%###########################################################################################
\section{Observations and data reduction}\label{secObs}

%%%%%%%%%%%%%%%%%%%%%%%%%%%%%%%%%%%%%%%%%%%%%%%%%%%%%%%%%%%%%%%%%%%%%%%%%%%%%%%%%%%%%%%%%%%%
\subsection{Interferometric observations with MIDI/VLTI}\label{secObsSubInt}

\begin{figure*}
     \centering
     \includegraphics[width=0.59\linewidth]{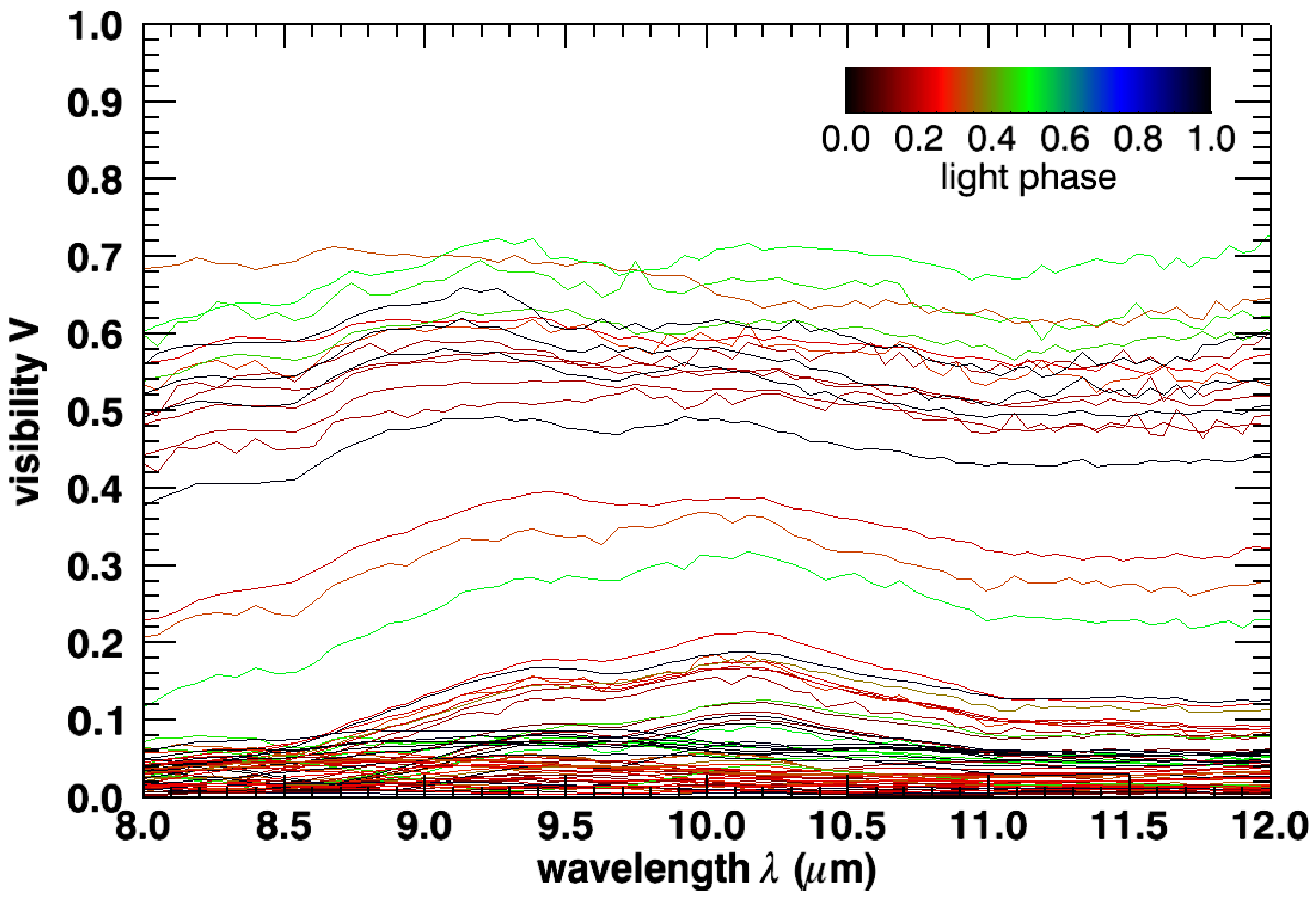}
     \includegraphics[width=0.40\linewidth]{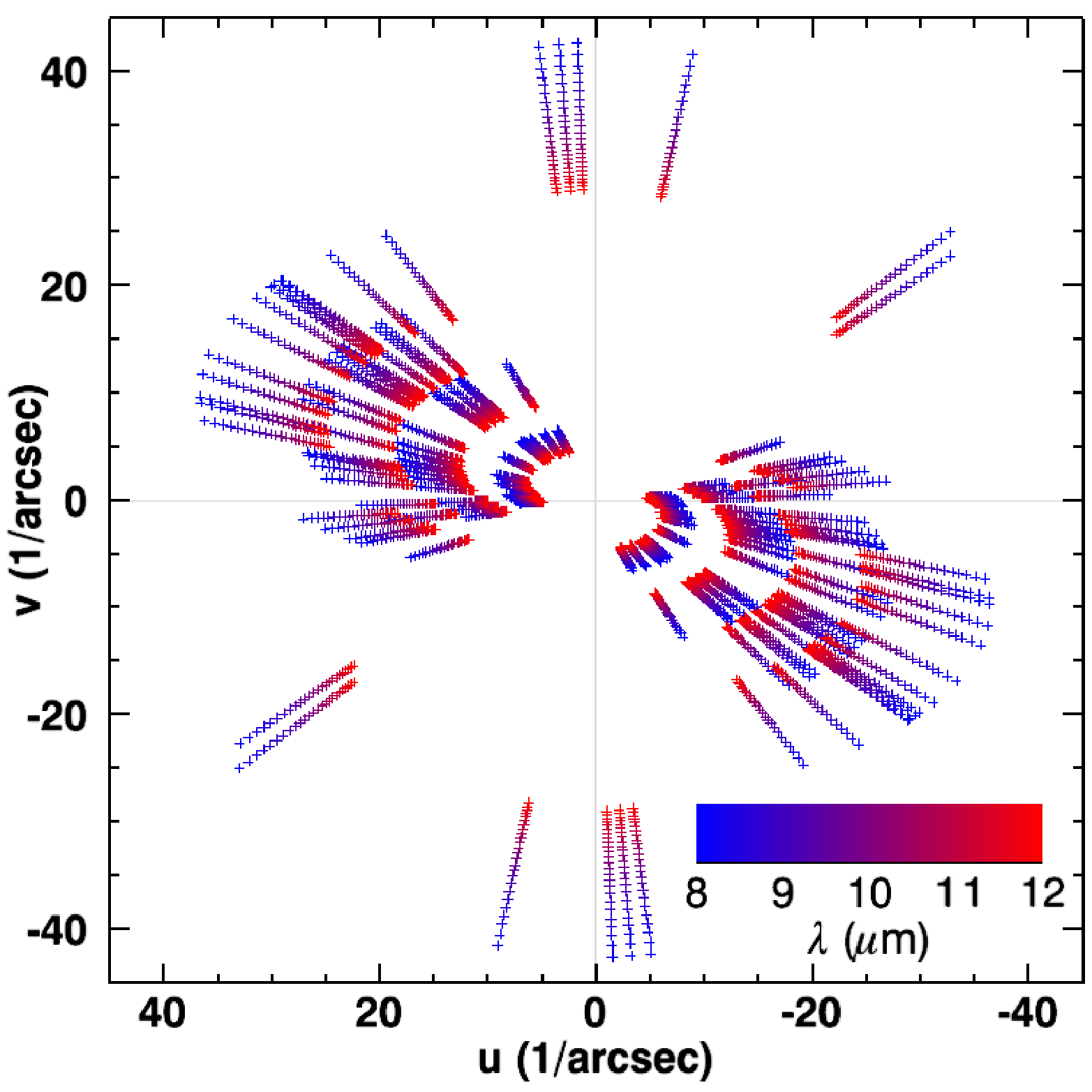}
     \caption{\textit{Left:} All 75 calibrated visibilities as function of wavelength obtained with MIDI, color-coded by visual light phase. Errors are omitted for clarity but given in Table~\ref{TableResults}. \textit{Right:} UV-coverage of all 75 used interferometric observations. The visibility spectra are binned into 25~wavelength bins.}
     \label{FigUVall}
   \end{figure*}

    The data presented here were obtained with the mid-IR ($8-13$~$\mu$m) interferometer MIDI (Leinert et al.~\cite{Leinert_et03}, \cite{Leinert_et04}) at the Very Large Telescope Interferometer (VLTI) in service mode. W~Hya is one of five stars monitored from P79 to P83 (April~2007 to September~2009) under the program IDs 079.D-0140, 080.D-0005, 081.D-0198, 082.D-0641, and 083.D-0294. An observation log is given in Table~\ref{TableObsLog}, showing all 83 individual observations. Six different baseline configurations of two auxiliary telescopes (ATs) were used. This results in projected baselines from 13 to 71 meter and a wide range of position angles (PA, $\vartheta$; east of north). The uv-coverage of all used data can be seen in the right hand panel of Fig.~\ref{FigUVall}.

%---------------------------------------------------------------
   \begin{table*}
     \caption{Observation log.}
     \label{TableObsLog}
     \centering
     \begin{tabular}{ccccccc|ccccccc}
       \hline
       \noalign{\smallskip}
        Date & AT$^{\mathrm{a}}$ & Disp$^{\mathrm{b}}$ & B$^{\mathrm{c}}$ ($\mathrm{m}$) & PA$^{\mathrm{d}}$ ($^\circ$) & Phase & QF$^{\mathrm{e}}$ & Date & AT$^{\mathrm{a}}$ & Disp$^{\mathrm{b}}$ & B$^{\mathrm{c}}$ ($\mathrm{m}$) & PA$^{\mathrm{d}}$ ($^\circ$) & Phase & QF$^{\mathrm{e}}$ \\
        \noalign{\smallskip}
        \hline
        \noalign{\smallskip}
	  2007-04-12 & A  & grism & 14.54 & 51.81 & 0.30 & used & 2008-04-03 & B & prism & 29.19 & 82.75 & 0.22 & used \\
	  2007-04-13 & B  & grism & 29.10 & 51.92 & 0.30 & used & 2008-04-28 & D & prism & 59.43 & 54.70 & 0.29 & used \\
	  2007-04-17 & B  & grism & 22.34 & 94.12 & 0.31 & used & 2008-04-28 & E & prism & 70.58 & 130.03& 0.29 & n.~u. \\
	  2007-04-22 & D  & grism & 59.53 & 54.92 & 0.33 & used & 2008-04-28 & E & prism & 71.32 & 10.41 & 0.29 & n.~u. \\
	  2007-04-22 & D  & grism & 59.53 & 54.92 & 0.33 & used & 2008-04-28 & D & prism & 45.23 & 93.72 & 0.29 & used \\
	  2007-04-22 & D  & grism & 42.06 & 96.19 & 0.33 & used & 2008-05-25 & D & prism & 61.21 & 58.90 & 0.36 & used \\
	  2007-04-22 & D  & grism & 38.56 & 99.06 & 0.33 & used & 2008-05-30 & D & prism & 30.02 & 107.46& 0.37 & used \\
	  2007-04-24 & E  & grism & 67.30 & 125.04& 0.33 & n.~u. & 2008-07-03 & A & prism & 13.67 & 86.31 & 0.46 & used \\
	  2007-04-25 & E  & grism & 64.92 & 123.16& 0.34 & n.~u. & 2008-07-03 & C & prism & 32.03 & 95.67 & 0.46 & used \\
	  2007-04-25 & F  & grism & 71.47 & 2.19  & 0.34 & used & 2008-07-06 & D & prism & 36.87 & 100.52 & 0.46 & used \\
	  2007-06-18 & A  & grism & 13.25 & 87.72 & 0.47 & used & 2009-01-16 & F & prism & 71.23 &$-12.23$& 0.96 & used \\
	  2007-06-20 & C  & grism & 46.98 & 62.82 & 0.48 & used & 2009-01-16 & D & prism & 60.22 & 56.51  & 0.96 & used \\
	  2007-06-20 & C  & grism & 43.73 & 82.84 & 0.48 & used & 2009-01-17 & E & prism & 65.19 & 123.34 & 0.97 & n.~u. \\
	  2007-07-02 & C  & grism & 48.00 & 70.64 & 0.51 & used & 2009-01-20 & C & prism & 46.78 & 61.99  & 0.97 & used \\
	  2007-07-02 & C  & grism & 40.49 & 86.86 & 0.51 & n.~u. & 2009-01-21 & B & prism & 25.33 & 32.67 & 0.98 & used \\
	  2007-07-04 & B  & grism & 31.55 & 75.51 & 0.52 & used & 2009-01-21 & B & prism & 28.19 & 47.79 & 0.98 & used \\
	  2007-07-04 & B  & grism & 24.39 & 90.99 & 0.52 & used & 2009-01-21 & B & prism & 29.88 & 55.45 & 0.98 & used \\
	  2008-01-10 & B  & prism & 30.08 & 56.39 & 0.01 & used & 2009-01-22 & C & prism & 43.10 & 50.26 & 0.98 & used \\
	  2008-02-20 & D  & grism & 59.95 & 55.89 & 0.11 & used & 2009-01-22 & C & prism & 44.68 & 55.03 & 0.98 & used \\
	  2008-02-20 & D  & prism & 62.76 & 63.21 & 0.11 & used & 2009-01-22 & A & prism & 45.94 & 59.00 & 0.98 & used \\
	  2008-02-21 & D  & prism & 63.89 & 68.30 & 0.11 & n.~u. & 2009-01-25 & A & prism & 12.34 & 28.13 & 0.99 & used \\
	  2008-02-22 & B* & prism & 31.85 & 73.49 & 0.12 & used & 2009-01-25 & A & prism & 13.05 & 37.34 & 0.99 & used \\
	  2008-02-22 & B* & prism & 31.00 & 78.00 & 0.12 & used & 2009-01-25 & A & prism & 14.35 & 50.06 & 0.99 & used \\
	  2008-03-02 & E  & prism & 66.93 & 124.69& 0.14 & used & 2009-01-25 & A & prism & 15.78 & 64.41 & 0.99 & used \\
	  2008-03-02 & F  & prism & 71.45 & 4.51  & 0.14 & used & 2009-01-27 & C & prism & 41.69 & 45.93 & 0.99 & used \\
	  2008-03-03 & D  & prism & 63.91 & 72.10 & 0.14 & used & 2009-01-27 & C & prism & 43.52 & 51.53 & 0.99 & used \\
	  2008-03-06 & D  & grism & 61.71 & 78.46 & 0.15 & used & 2009-02-16 & D & prism & 63.97 & 69.21 & 0.04 & used \\
	  2008-03-11 & A  & grism & 12.33 & 28.02 & 0.16 & used & 2009-02-16 & D & prism & 63.24 & 75.18 & 0.04 & used \\
	  2008-03-11 & A  & prism & 13.98 & 46.70 & 0.16 & used & 2009-03-16 & D & prism & 62.93 & 76.01 & 0.12 & used \\
	  2008-03-12 & A  & grism & 15.23 & 79.71 & 0.17 & used & 2009-04-20 & B & prism & 20.28 & 97.36 & 0.21 & used \\
	  2008-03-13 & B  & grism & 30.54 & 79.42 & 0.17 & used & 2009-04-23 & C & prism & 45.26 & 80.46 & 0.21 & used \\
	  2008-03-13 & B  & prism & 31.94 & 68.34 & 0.17 & used & 2009-04-23 & C & prism & 36.25 & 91.34 & 0.21 & used \\
	  2008-03-13 & B  & prism & 29.95 & 81.01 & 0.17 & used & 2009-04-24 & B & prism & 27.75 & 85.53 & 0.22 & used \\
	  2008-03-14 & A  & prism & 15.86 & 65.67 & 0.17 & n.~u. & 2009-05-02 & D & prism & 55.99 & 46.91 & 0.24 & used \\
	  2008-03-14 & A  & prism & 15.80 & 75.37 & 0.17 & used & 2009-05-03 & E & prism & 69.19 & 127.28& 0.24 & used \\
	  2008-03-14 & A  & prism & 15.26 & 79.55 & 0.17 & used & 2009-05-03 & F & prism & 71.42 & 7.00  & 0.24 & used \\
	  2008-03-25 & C  & prism & 47.86 & 67.68 & 0.20 & used & 2009-05-03 & D & prism & 52.46 & 38.03 & 0.24 & used \\
	  2008-03-25 & C  & prism & 47.38 & 75.34 & 0.20 & used & 2009-06-04 & C & prism & 45.75 & 58.37 & 0.32 & used \\
	  2008-03-25 & C  & prism & 41.85 & 85.29 & 0.20 & used & 2009-06-04 & C & prism & 46.29 & 60.19 & 0.32 & used \\
	  2008-04-01 & B  & prism & 28.32 & 48.37 & 0.22 & used & 2009-06-04 & A & prism & 12.40 & 90.40 & 0.32 & used \\
	  2008-04-02 & C  & prism & 44.91 & 81.06 & 0.22 & used & 2009-08-15 & A & prism & 11.73 & 92.45 & 0.51 & used \\
	  2008-04-02 & A  & prism & 12.96 & 88.65 & 0.22 & used &  &  &  &  &  &  &  \\
        \noalign{\smallskip}
        \hline
      \end{tabular}
    \begin{flushleft}
       $^{\mathrm{a}}$~AT stations: A~=~E0$-$G0, B~=~G0$-$H0, B*~=~A0$-$D0, C~=~E0$-$H0, D~=~D0$-$H0, E~=~D0$-$G1 and F~=~H0$-$G1; 
       $^{\mathrm{b}}$~dispersive element; 
       $^{\mathrm{c}}$~projected baseline length; 
       $^{\mathrm{d}}$~position angle of the projected baseline on the sky; 
       $^{\mathrm{e}}$~quality flag showing if that observation is used for the model fitting or not (n.~u.), see Sect.~\ref{secModSubFDD} for reasons that a value had not been used
     \end{flushleft}
   \end{table*}
%---------------------------------------------------------------
%---------------------------------------------------------------
%---------------------------------------------------------------
    
    Before or after each target observation the calibrator 2~Cen (\object{HD~120323}) was observed with the same setup to calibrate the visibility measurements. The calibrator star has a diameter of (13.25~$\pm$~0.06)~mas (model diameter from Verhoelst~\cite{Verhoelst_05}\footnote{http://www.ster.kuleuven.ac.be/$\sim$tijl/MIDI\_calibration/mcc.txt;~see also ESO CalVin database: http://www.eso.org/observing/etc/}), a spectral type of M4.5~III, a 12~$\mu$m flux of (256~$\pm$~26)~Jy (IRAS\footnote{http://irsa.ipac.caltech.edu/Missions/iras.html}), and an angular separation to the target of about 6$^\circ$. Observations were executed in \texttt{SCI-PHOT} mode, where the photometric and the interferometric spectra are recorded simultaneously. This has the advantage that the photometry and the fringe signal are observed under the same atmospheric conditions. Either the prism, with a spectral resolution of $R = \lambda/\Delta\lambda = 30$, or the grism, with a spectral resolution of 230, were used to obtain spectrally dispersed fringes.

%%%%%%%%%%%%%%%%%%%%%%%%%%%%%%%%%%%%%%%%%%%%%%%%%%%%%%%%%%%%%%%%%%%%%%%%%%%%%%%%%%%%%%%%%%%%
\subsection{MIDI Sci-Phot data reduction}\label{secObsSubRed}

\begin{figure}
     \centering
     \includegraphics[width=0.95\linewidth]{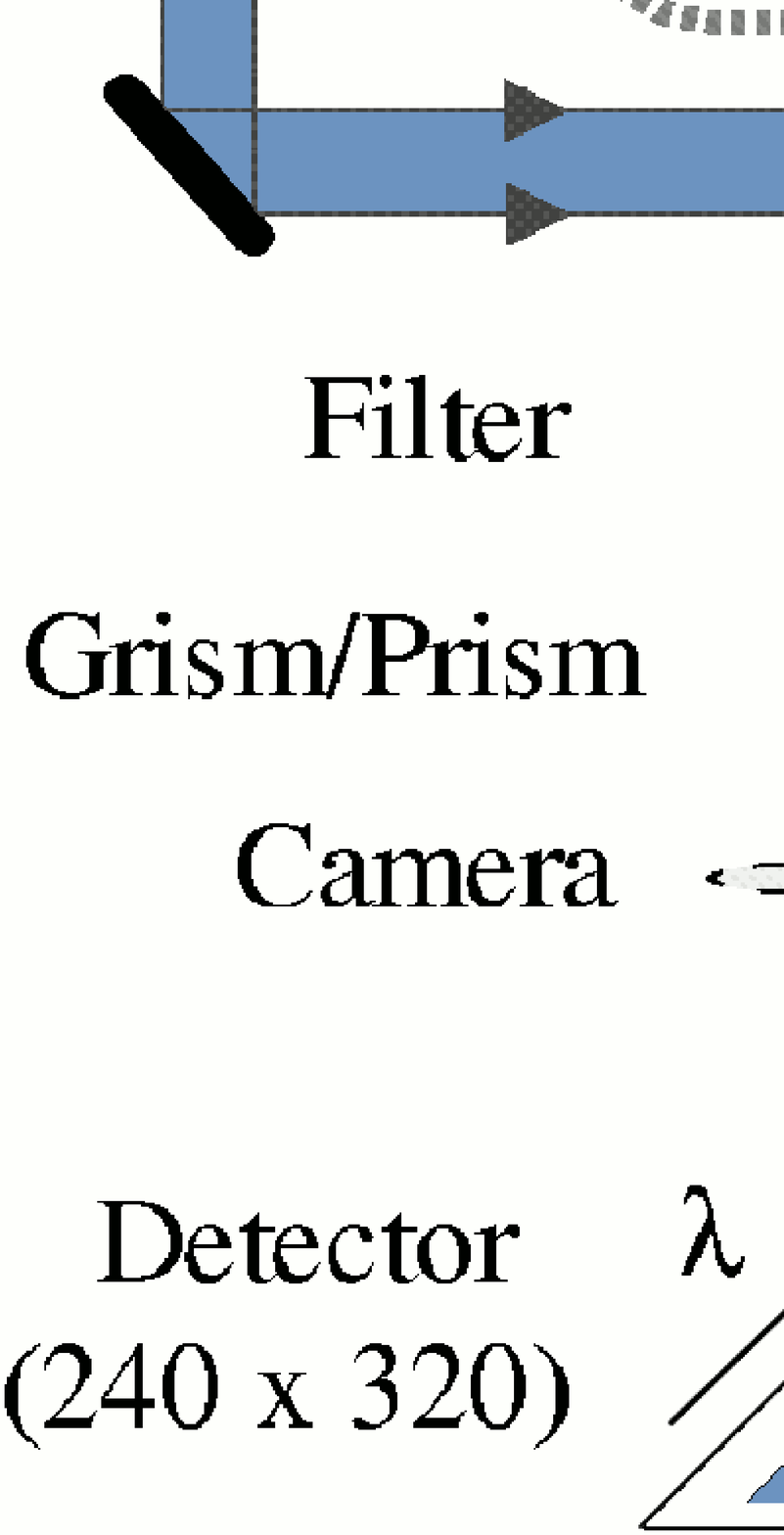}
     \caption{MIDI setup in \texttt{SCI-PHOT} mode (adopted from Leinert et al.~\cite{Leinert_et03}).}
     \label{FigMidi}
   \end{figure}

    The standard \texttt{MIA+EWS}\footnote{http://www.strw.leidenuniv.nl/$\sim$koehler/MIDI} (version~1.6) data reduction package with additional routines for processing \texttt{SCI-PHOT} data (Walter Jaffe, private communication) was used. In any observation in \texttt{SCI-PHOT} mode, all four read-out windows (channels) are illuminated simultaneously. The two central channels record the interferometric signals, while the two outer channels are dedicated to measuring photometric fluxes (cf.~Fig.~\ref{FigMidi}). To measure these signals, the two telescope beams (one beam from each telescope) are separated by a beamsplitter. One of the separated beams of each telescope is sent directly to the photometric detector window, while the other two remaining beams (one from each telescope) are combined in the beam combiner to obtain two interferometric signals with opposite phases.

%---------------------------------------------------------------
%---------------------------------------------------------------

    To be able to calculate the instrumental visibility by dividing the interferometric fringe signal (correlated flux) by the photometric signal (uncorrelated flux), the exact light split ratios between these channels have to be known as functions of wavelength. This requires recording \emph{additional photometry}. By closing the shutter of one telescope, the light from the other telescope illuminates one of the photometric channels and both interferometric channels. From these data, the light split ratio between photometric and interferometric channels for this telescope beam can be computed. This is repeated for the other telescope beam as well. In principle, this would only be necessary once per night or less often, but is done for every observation to check for consistency.

    Accordingly, the first step in the reduction is the calculation of these ratios (or better the mapping behavior between interferometric and photometric channels). After unchopping the \emph{additional photometry} data, a point spread function is fitted perpendicular to the wavelength direction as a function of wavelength to each of the three read-out windows to obtain the spectra. This step also includes the determination of the sky and instrumental background, remaining after the unchopping, and its removal. The spectra of all channels are then divided by each other to derive the respective split ratios. One of the largest error sources here is the determination of the correct background, since it can result in too high or too low fluxes in affected channels and can therefore lead to errors in the final visibility and photometric spectra. Typical final errors are on the order of 10\% and 30\%, respectively.

    With the obtained light split ratios (mapping behavior), the \texttt{SCI-PHOT} data are reduced in the next step. After unchopping the \texttt{SCI-PHOT} data, the simultaneously recorded photometric spectra (in the photometric channels) are multiplied by the appropriate split ratios; i.e.,~each photometric spectrum is mapped to each interferometric channel to get an estimate of the uncorrelated (photometric) fluxes at the location of the interferometric channels. The average of the summed images of each channel gives the final raw photometry. To obtain the visibility, the interferometric fringe signal (correlated flux) in each interferometric channel is divided by the estimated geometric mean image of the mapped uncorrelated (photometric) spectra. Prior to the division, the fringes are coherently averaged after determining the group delay, similar to the \texttt{HIGH-SENS} reduction (cf.~e.g.~Ratzka \cite{Ratzka05}). Both visibility spectra are then averaged to get the final raw visibility. A more detailed description can be found in Zhao-Geisler (\cite{ZhGeisler10}).
    
    These steps are taken separately for the target (W~Hya) and the calibrator (2~Cen). With the known calibrator diameter the raw target visibility is calibrated with the calibrator's transfer function, assuming a wavelength-independent uniform disk diameter. The raw photometry is calibrated by assuming that the spectrum of the calibrator looks close enough to the Rayleigh-Jeans part of a blackbody and by using the known 12~$\mu$m flux of the calibrator. The Rayleigh-Jeans assumption is not fully valid between 8 and 9~$\mu$m since 2~Cen shows a weak SiO absorption feature in its ISO spectrum.

    After reducing and calibrating all data, results with unphysical visibilities (due to bad environmental conditions or failure of the reduction process) are rejected. Eight of the 83 observations were thus excluded and are flagged as ``not used'' in Table~\ref{TableObsLog}. Since several observations of W~Hya were executed in some nights, more than one calibrator observation was available. For those nights, the calibrators nearest in time were used to calculate a median calibrated visibility with the standard deviation as a typical error.

    Grism data were interpolated to the prism grid. The wavelength range between 8 and 12~$\mu$m was binned into 25 wavelength bins to allow a faster computation of the model fits. Within each wavelength bin a mean visibility and error were calculated. For the fits later on, we always checked that this approach does not mask any additional spectral features, with the result that in all cases the shapes of the wavelength-dependent parameters were not altered significantly. Measurements beyond 12~$\mu$m were not used because the recorded flux was too low, making the reduction inconsistent.
    
    To avoid the problem of underestimating the error and to allocate an error to visibilities where only one calibrator observation exists, all available visibility errors within the same wavelength bin were averaged. The resulting mean error was then assigned to \emph{all} visibilities within this wavelength bin. The final 75 visibility curves are shown in the left hand panel of Fig.~\ref{FigUVall}. The central wavelengths and assigned visibility errors of each bin are given in the first and second columns of Table~\ref{TableResults}, respectively.
    
    Assuming the same error for all measurements within a wavelength bin improves the model fits as well. In a fitting method based on a chi-square technique (cf.~Sect.~\ref{secModSubFDD}), all visibility measurements are then represented with the same significance. In the case that each value has an individual uncertainty, values with higher errors are underrepresented, while values with low errors are overrepresented. Consequently, higher visibilities, with in general higher absolute errors, are not weighted accordingly, and the importance of lower visibilities, at generally higher spatial frequencies with lower absolute errors, is overestimated. Only equal weighting ensures that over the whole spatial frequency range a model fits all data well, assuming that all measurements are equally significant. This has the consequence that chi-square values are only useful in a relative sense. For this reason chi-square values are not given.

%%%%%%%%%%%%%%%%%%%%%%%%%%%%%%%%%%%%%%%%%%%%%%%%%%%%%%%%%%%%%%%%%%%%%%%%%%%%%%%%%%%%%%%%%%%%
\subsection{Light curves}\label{secObsSubLC}

%---------------------------------------------------------------
   \begin{figure*}
     \centering
     \includegraphics[width=0.95\linewidth]{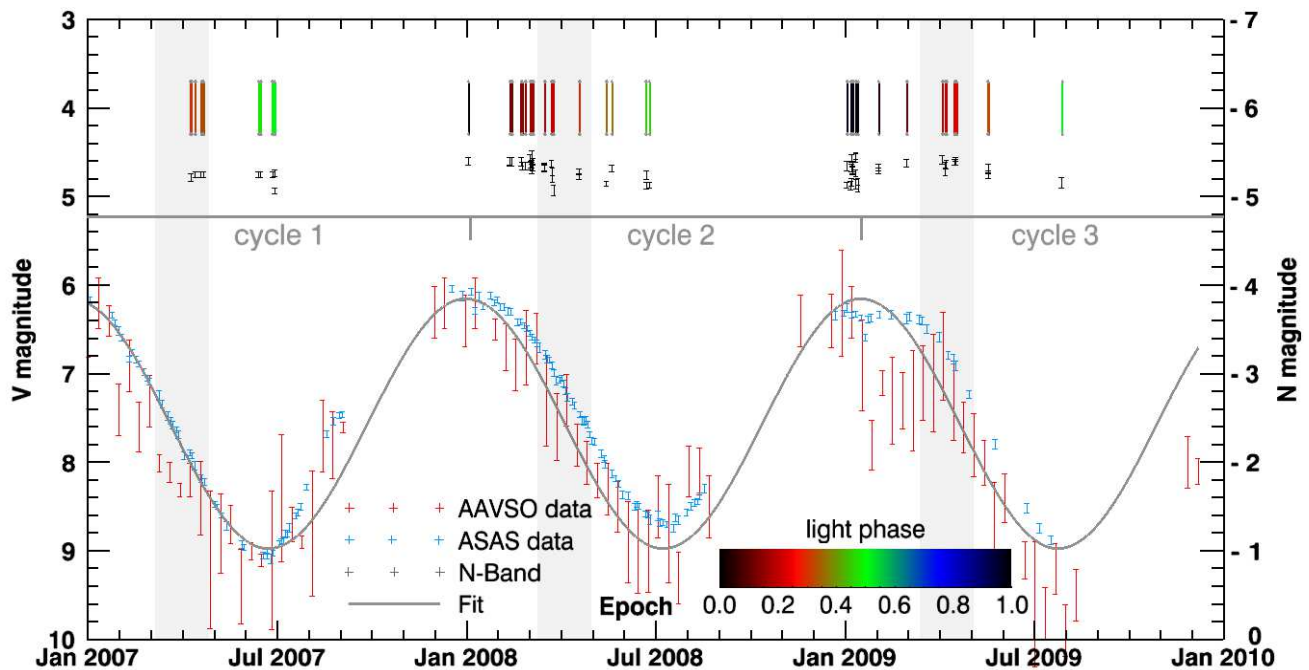}
     \caption{The visual light curve of W~Hya, covering the period of the MIDI observations. The V magnitudes are from AAVSO (10~day bins) and ASAS. A simple sinusoidal fit is included to determine the phases used throughout this paper. The times of the MIDI observations are included as tick marks above the curve and are color-coded by visual light phase. The MIDI flux at around 12~$\mu\mathrm{m}$ is shown directly below the tick marks with the magnitude scale given on the right. The data within the three shaded regions are used for cycle-to-cycle studies in Sect.~\ref{secPhaseSubCycle}.}
     \label{FigLight_t}
   \end{figure*}
%---------------------------------------------------------------
%---------------------------------------------------------------
   \begin{figure*}
     \centering
     \includegraphics[width=0.95\linewidth]{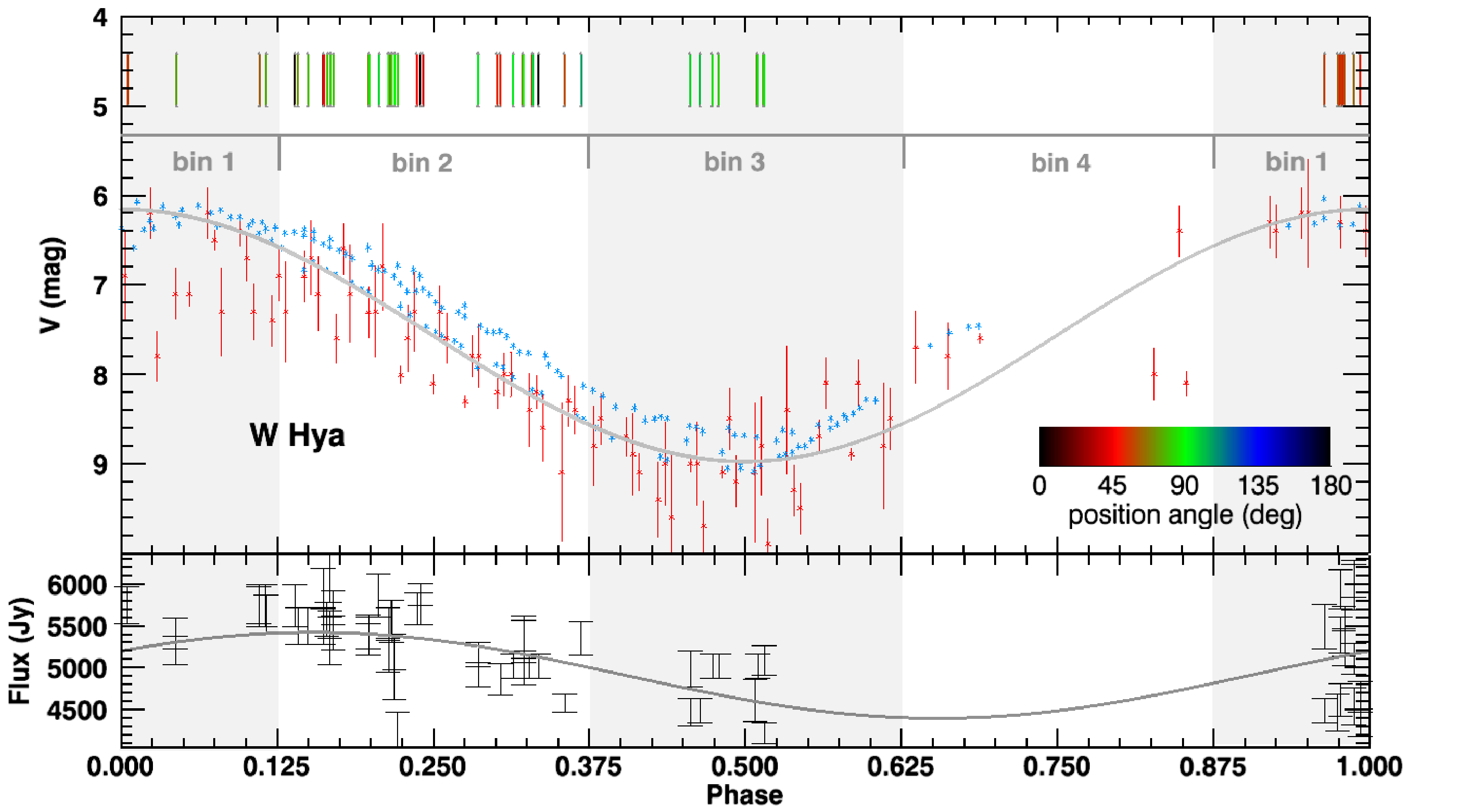}
     \caption{Same as Fig.~\ref{FigLight_t}, but plotted versus visual light phase. The MIDI observations are color-coded by position angle. The shaded and white regions (bins 1~to~4) are used for intracycle studies in Sect.~\ref{secPhaseSubLight}.}
     \label{FigLight_p}
   \end{figure*}
%---------------------------------------------------------------

    To assign a pulsation phase to the observations, visual data from the American Association of Variable Star Observers (AAVSO\footnote{http://www.aavso.org/}) and the All Sky Automated Survey (ASAS\footnote{http://www.astrouw.edu.pl/asas/}; Pojmanski et al.~\cite{Pojmanski_et2005}) are used. After binning the AAVSO data into ten-day bins, a simple sinusoid was fitted to the AAVSO and ASAS data over a period of about 10~years ($2000-2010$). The fit gives a period of (388~$\pm$~5)~days, a mean visual magnitude of (7.6~$\pm$~0.1)~mag, a semi-amplitude of (1.4~$\pm$~0.1)~mag and a Julian Date of maximum brightness of (2452922~$\pm$~5)~days (defined as phase 0.0). These data, including the fit, are plotted in Fig.~\ref{FigLight_t} versus time and in Fig.~\ref{FigLight_p} versus visual light phase. The visual amplitude of the semi-regular variable W~Hya is much lower than typical Mira variables.

    In both plots, the times of the 75 used MIDI observations are included as tick marks above the light curve, color-coded by phase in Fig.~\ref{FigLight_t} and color-coded by position angle in Fig.~\ref{FigLight_p}. This allows recognizing different distributions of position angles throughout the pulsation cycle. It can be seen that the position angles are not sampled well over the pulsation period. At around maximum light, position angles cluster around 50$^\circ$, while around minimum light, position angles are mostly at around 90$^\circ$. Only for the intermediate phase is the position angle sampling more uniform. This will be important in Sect.~\ref{secPhase}, where asymmetries, intracycle variations and cycle-to-cycle variations are investigated. The regions used for this study are already shaded in both figures. It is also notable that the whole pulsation cycle is not equally sampled because the pulsation period is around one year (also true for the AAVSO and ASAS sampling).

    The MIDI photometry is shown in Fig.~\ref{FigLight_t} directly below the tick marks with the magnitude scale given on the right and as flux values below the magnitude plot in Fig.~\ref{FigLight_p}. Only the averaged fluxes between 11.5 and 12.5~$\mu$m are plotted in order to be comparable with observations made with other instruments (e.g.~IRAS). Even though the MIDI fluxes are afflicted by fairly large errors, a clear phase dependence is detectable in the phase-folded plot. A sinusoidal fit gives a semi-amplitude of (510~$\pm$~100)~Jy, which is equivalent to a flux variation on the order of 20\% between maximum and minimum light, with a mean mid-IR flux of (4.9~$\pm$~0.2)~kJy. The maximum occurs after the visual maximum at visual phase 0.15~$\pm$~0.05.

    This phase shift is consistent with previous studies of AGB stars (cf.~e.g.~ Lattanzio \& Wood \cite{Lattanzio_Wood_04}, Smith et al.~\cite{Smith_et06} and references therein, and Nowotny et al.~\cite{Nowotny_et10}). The reasons for this are that most of the flux is emitted in the infrared and that the visual spectrum is strongly influenced by features of the temperature-sensitive TiO molecule (Nowotny et al.~\cite{Nowotny_et10}). Similar flux variations have been reported for other AGB stars by Ohnaka (private communication; preliminary results for R~Car, R~Cnc and Z~Pup) for MIDI AT observations, by Wittkowski et al.~(\cite{Wittkowski_et07}) (O-rich Mira S~Ori), and Ohnaka et al.~(\cite{Ohnaka_et07}) (C-rich Mira V~Oph) for MIDI Unit Telescope (UT) observations, and by Karovicova (private communication; preliminary result for RR~Aql) for both AT and UT observations. Since the photometric accuracy of MIDI used with the ATs is very low, these values should be used with caution. For this reason all data are interpreted with respect to the visual light curve instead of the mid-IR light curve. It should also be kept in mind that folding consecutive cycles might not always be appropriate since the pulsation is not strictly regular. The irregularity can clearly be inferred from the plots.

%%%%%%%%%%%%%%%%%%%%%%%%%%%%%%%%%%%%%%%%%%%%%%%%%%%%%%%%%%%%%%%%%%%%%%%%%%%%%%%%%%%%%%%%%%%%
\subsection{Spectra}\label{secObsSubSpec}

\begin{figure*}
     \centering
     \includegraphics[width=0.95\linewidth]{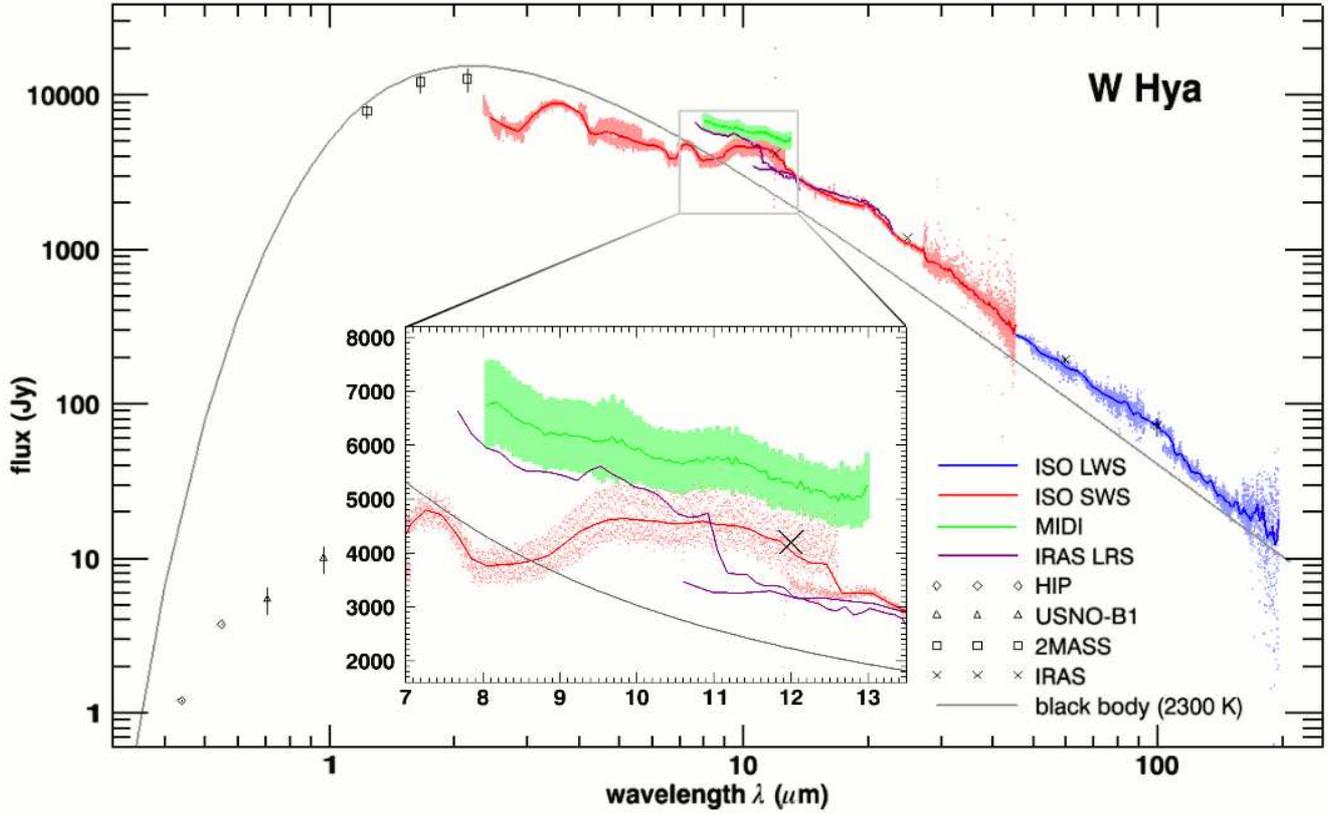}
     \caption{Spectral energy distribution of W~Hya. The included Photometry is not corrected for reddening effects and the black body curve is only included for guidance. The ISO LWS and SWS spectra from observation epochs 1996-02-07 and 1992-02-14 are used, respectively. The MIDI spectrum (green, where light green are the errors) does not show the silicate emission feature at around 10~microns. The MIDI spectrum is shown in the inset in more detail.}
    \label{FigSpec}
   \end{figure*}

    The median of all calibrated photometric MIDI spectra (scaled with the 12~$\mu$m flux of the calibrator as described above) is shown in the spectral energy distribution (SED) in Fig.~\ref{FigSpec}. The uncertainties are given by the standard deviation. Even if this averaging over all pulsation phases and cycles might be questionable, it is not important here, since the spectrum is not used for any modeling. In addition to fully scientifically verified spectra from ISO\footnote{http://iso.esac.esa.int/ida/} and IRAS, photometric data from HIPPARCOS, USNO-B1, 2MASS, and IRAS\footnote{all from http://vizier.u-strasbg.fr/viz-bin/VizieR}, not corrected for reddening effects, are plotted as well. A blackbody with an effective temperature of 2300~K is overplotted as guidance. Due to the infrared excess, strong metallic oxide line, and molecule absorption at shorter wavelengths, and dust extinction, it is not expected that a blackbody curve fits the spectral data in an appropriate way.
    
    The ISO SWS spectrum is mainly dominated by absorption bands of H$_2$O between $2.5-3.0$~$\mu$m (stretching mode) and $5.0-8.0$~$\mu$m (bending mode), and an SiO absorption band between 8 and~9~$\mu$m ($\nu$~=~$1-0$). Distinct absorption lines of CO at around 2.4~$\mu$m, OH at $2.9-4.0$~$\mu$m, CO$_2$ at 4.25~$\mu$m and SO$_2$ at 7.4~$\mu$m can be weakly seen in the spectrum as well. Justtanont et al.~(\cite{Justtanont_et04}) derived from temperature investigations that these absorptions originate in different molecular layers in the circumstellar shell. The OH, CO, and CO$_2$ absorption bands arise mainly from a hot (about 3000~K), dense region very close to the stellar photosphere, where H$_2$O is still photodissociated by shocks. The H$_2$O and a second CO$_2$ absorption band originate in a layer with a temperature of 1000~K, i.e.~a molecular layer (molsphere\footnote{The occurrence of molecular layers is the result of the capability of forming certain molecules at a specific distance from the star and their dilution when reaching larger distances. Such a shell exists for all atmospheric molecules, and only their total abundance and radiative properties determine whether such a molsphere can be detected or not.}) farther out. The SiO molecule absorption arises in the same region where the H$_2$O molecular shell exists and where SiO is still not condensed in dust grains. %Justtanont et al.~(\cite{Justtanont_et05}): a relatively high H$_2$O abundance of (2.0~$\pm$~1.0)~$\times$~$10^{-3}$.

    Dust emission can also be identified in the ISO spectra. The features between 10 and 20~$\mu$m are a combination of emission from amorphous silicates at around 10~$\mu$m, compact Al$_2$O$_3$ at around 11~$\mu$m, and MgFeO at around 19~$\mu$m. Justtanont et al.~(\cite{Justtanont_et04}) have obtained a satisfactory SED fit to all three emission features. They derived a low\footnote{If compared with a typical AGB star.} total mass-loss rate of (3.5$-$8)~$\times$~$10^{-8} M_{\odot}$yr$^{-1}$, and a lower limit to the dust mass-loss rate for silicates, Al$_2$O$_3$ and MgFeO of 1.5~$\times$~$10^{-10} M_{\odot}$yr$^{-1}$, 1.3~$\times$~$10^{-10} M_{\odot}$yr$^{-1}$ and 2.5~$\times$~$10^{-11} M_{\odot}$yr$^{-1}$ respectively. Hinkle et al.~(\cite{Hinkle_et97}) and de Beck at al.~(\cite{deBeck_et10}) obtained similar total mass-loss rates of 2~$\times$~$10^{-8} M_{\odot}$yr$^{-1}$ and 7.8~$\times$~$10^{-8} M_{\odot}$yr$^{-1}$, respectively. The derived low dust mass-loss rate and pulsation behavior is typical of a star at the beginning of the AGB phase and the transformation from a semi-regular variable to a Mira star (Hinkle et al.~\cite{Hinkle_et97}). W~Hya is classified in the dust emission scheme of Sloan and Price (\cite{SloanPrice_98}) as SE8 (classic narrow silicate emission type).

    If the MIDI spectrum is compared with the ISO SWS spectrum, it is obvious that the silicate emission is not detected. This is shown in more detail in the inset of Fig.~\ref{FigSpec}. In the MIDI spectrum of W~Hya, a weak emission of amorphous Al$_2$O$_3$ at around 11~$\mu$m is present, while no silicate feature at around 10~$\mu$m can be seen\footnote{Fluctuations between 9.3 and 10~$\mu$m are caused by difficulties in the reduction caused by the telluric ozone feature, and flux measurements beyond 12~$\mu$m are probably not calibrated well.}. This behavior can be attributed to instrumental characteristics. ISO has a much larger field of view (FOV) compared to MIDI. With a small FOV of about 1 to 2~arcsec\footnote{The exact value depends on the used AT baseline, the slit/mask position and other instrumental specifications.}, the emission of the extended silicate dust shell is not detected in the MIDI spectrum.
    
    However, this non-detection allows a lower limit to be derived for the inner boundary of the silicate dust shell. Assuming a conservative value of the FOV of 1~arcsec, the main emission of the silicate dust shell in W~Hya originates in a region with an inner radius larger than 28 photospheric radii\footnote{The value of the photospheric radius, $R_{\mathrm{phot}} = \theta_{\mathrm{phot}}$/2, is defined in Sect.~\ref{secPhaseSubWave}} ($>$50~AU, $>$0.5~arcsec). This is consistent with knowing that the dust envelope is very extended, up to 40~arcsec (Hawkins \cite{Hawkins90}). The higher flux measured with MIDI compared to ISO can be attributed to the circumstance that the averaged spectrum is dominated by observations conducted around mid-IR maximum phases, while the ISO spectrum were obtained in a post visual minimum phase (phase~0.8, which is close to the mid-IR minimum; cf.~Fig.~\ref{FigLight_p}). Together with the calibration uncertainty of the averaged MIDI spectrum of larger than 10\%, this apparent discrepancy can be understood.

%###########################################################################################
%###########################################################################################
\section{Modeling the visibility data}\label{secMod}

    The most straight forward way of interpreting sparsely sampled interferometric data (visibilities) is by fitting the Fourier transform of an assumed brightness distribution of the object. Simple size estimations can be obtained by elementary models with only a few free parameters. As mentioned in the introduction, the definition of a diameter is difficult because of its strong wavelength dependence, as well as of the intracycle and possible long-term cycle-to-cycle variations (e.g.~Haniff et al.~\cite{Haniff_et95}) expected from models (e.g.~Hofmann et al.~\cite{Hofmann_et98}). On the other hand, the size and its wavelength-dependent shape can tell which layer of the atmosphere is actually observed and which chemical and physical mechanism are responsible for this appearance. The low surface gravity results in an extended atmosphere and temperature structure and in the formation of molecular layers around late-type stars. Therefore, no sharp transition between star and circumstellar environment exists and limb-darkening effects (center-to-limb variations) are very pronounced.

\begin{figure*}
     \centering
     \includegraphics[width=0.49\linewidth]{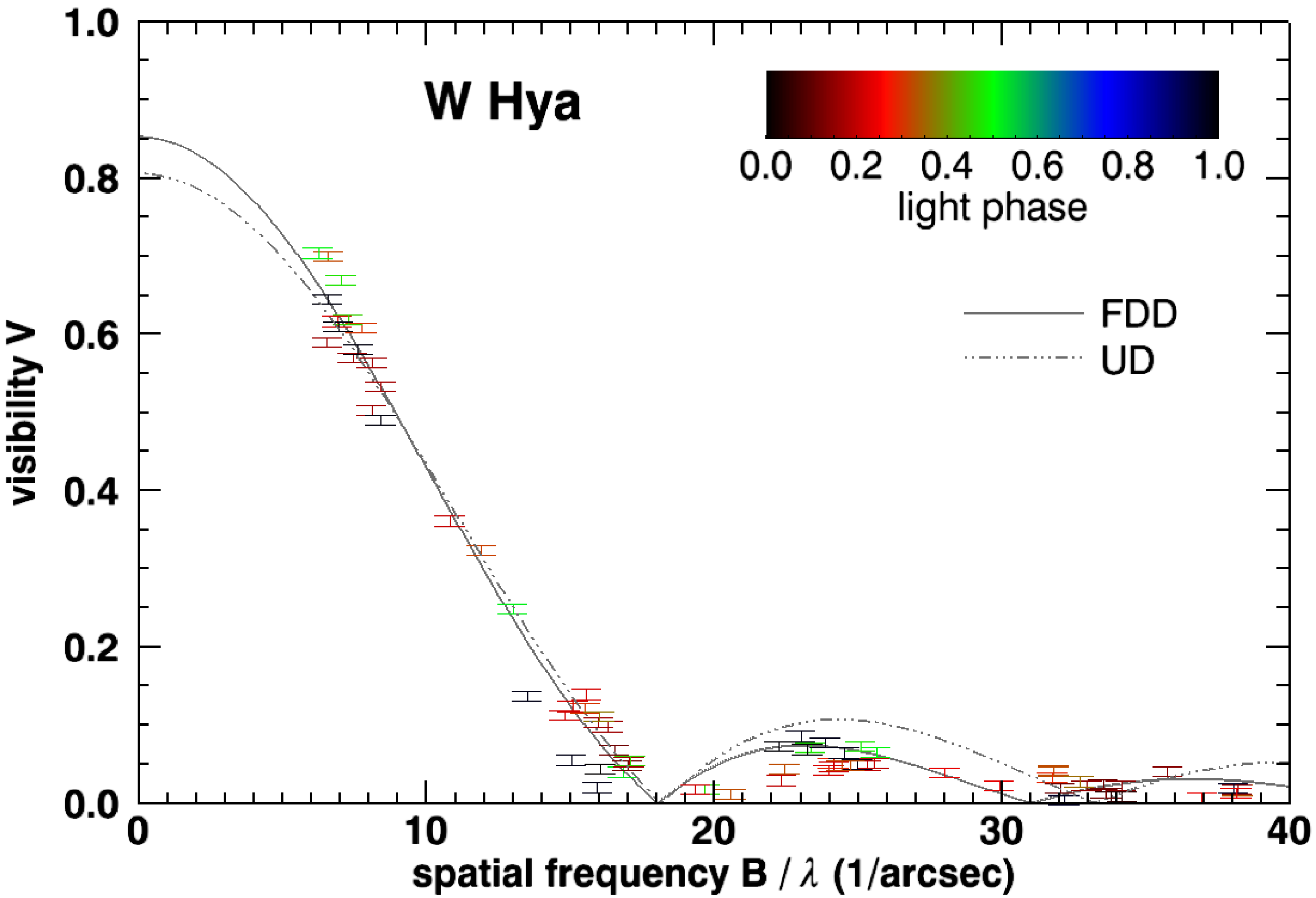}
     \includegraphics[width=0.49\linewidth]{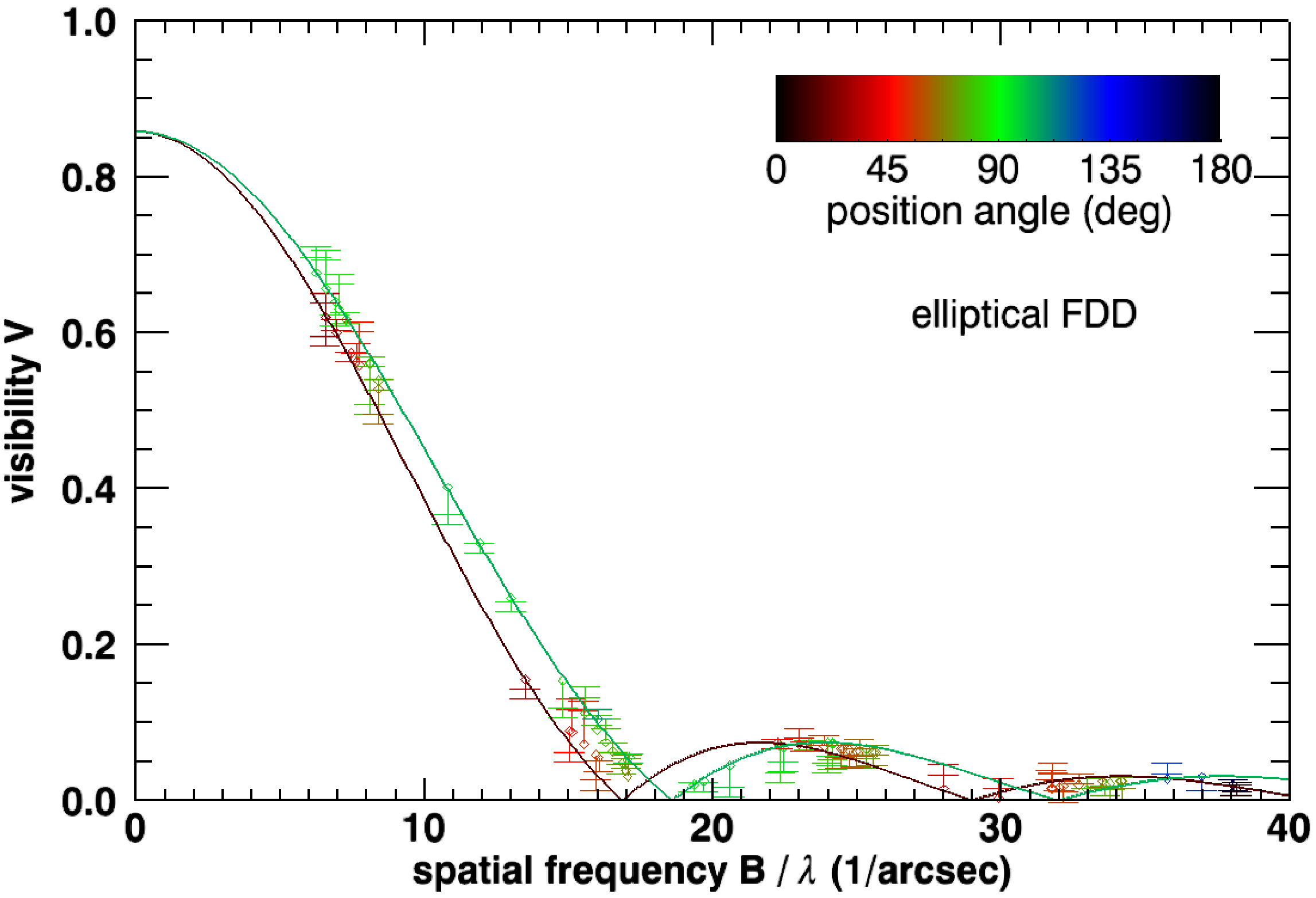}
     \caption{Fits of different models to the visibility measurements for wavelength bin~6 (9.07~$\mu$m). The 75 visibility data, taken over all pulsation phases and cycles, are plotted versus spatial frequency. \textit{Left:} The measurements color-coded by visual light phase. The fit of a circular FDD gives for this wavelength a diameter of $79.2$~mas and a relative flux contribution of $0.85$. In addition, it can be clearly seen that a UD does not fit the data as well as an FDD. \textit{Right:} The measurements color-coded by position angle. The elliptical FDD model fit results along the minor (76~mas) and major (85~mas) axis are included as solid lines. The fit to a given measurement is represented as a small open circle and is connected by a line with the corresponding observational point. The relative flux contribution is determined to be $0.86$.}
     \label{FigSymModel}
   \end{figure*}

    All 75~visibility measurements of W~Hya are plotted in Fig.~\ref{FigSymModel} for one example wavelength bin (9.07~$\mu$m) as function of spatial frequency (projected baseline divided by wavelength). There is a considerable spread in the data, but this should be expected since the measurements were obtained at different pulsation phases, pulsation cycles and position angles. In addition, some uncertainties in the reduction process remain. The left hand panel shows the data color-coded by visual light phase, while the right hand panel shows the data color-coded by position angle. In both plots systematically displaced distributions are noticeable. In the left hand panel, observations at visual maximum (dark points) have lower visibilities than those at pre-minimum (red points), if measurements with similar spatial frequencies are compared. However, this trend is very weak. In comparison, the displacement of the two distributions with different position angles is more significant in the right hand panel. For a discussion of this see Sect.~\ref{secPhase}.

    After the first zero, the visibilities in the second lobe at around 24~arcsec$^{-1}$ remain low. This indicates that a uniform disk (UD), with a constant brightness distribution up to the edge of the disk, cannot be applied to the data. A UD gives a higher second lobe in Fourier space. On the other hand, a simple Gaussian is not appropriate either, since a second lobe is clearly detected. Only a model that takes an extended atmosphere into account, which results in limb-darkening, can fit the data properly. The simplest possibility with only a few free parameters and a low second lobe is a fully limb-darkened disk (FDD). To account for a flux contribution of the extended silicate dust shell, the visibility function is not forced to be 1 at zero spatial frequency. Due to the lack of measurements at very low spatial frequencies, the dimension of the silicate dust shell cannot be constrained interferometrically.

%---------------------------------------------------------------
%---------------------------------------------------------------

%---------------------------------------------------------------
%---------------------------------------------------------------

%%%%%%%%%%%%%%%%%%%%%%%%%%%%%%%%%%%%%%%%%%%%%%%%%%%%%%%%%%%%%%%%%%%%%%%%%%%%%%%%%%%%%%%%%%%%
\subsection{Fully limb-darkened disk (FDD)}\label{secModSubFDD}
    
    A fully limb-darkened disk model is fitted to the visibility measurements in order to get diameter estimates. The visibility and intensity functions of a circular FDD are given by
    \begin{eqnarray}
      V \; (r)&=& \left|\,\epsilon\;\;\frac{3\sqrt{\pi}\;J_{3/2}(\pi\theta_{\mathrm{FDD}} r)}{\sqrt{2}\;(\pi\theta_{\mathrm{FDD}} r)^{3/2}}\,\right|\;\mbox{and}\\[4mm]
      I \; (\rho) &=& \left\{ \begin{array}{ll}
         \epsilon\;\;\frac{8}{\pi^2 \theta^2_{\mathrm{FDD}}}\sqrt{1-\left(\frac{2\rho}{\theta_{\mathrm{FDD}}}\right)^2} & \;\mbox{if} \;\rho \leq \frac{\theta}{2}\\[3mm]
         0 & \;\mbox{otherwise,} \end{array} \right.
    \end{eqnarray}
    \noindent respectively, where $J_{3/2}$ is a Bessel function of the first kind of order 3/2, $\theta_{\mathrm{FDD}}$~the FDD diameter and $\epsilon$~the flux contribution of the FDD. The flux contribution $\epsilon$ can be less than~1 if a fully resolved component (FRC), e.g.~a surrounding silicate dust shell, adds flux at the given wavelength, since the total visibility consists of both components:
    \begin{eqnarray}
      V_{\mathrm{total}} &=& |\,\epsilon \; \mathcal{V}_{\mathrm{FDD}} + (1-\epsilon) \; \mathcal{V}_{\mathrm{FRC}}\,|
    \end{eqnarray}
    \noindent The complex visibility of a fully resolved component, $\mathcal{V}_{\mathrm{FRC}}$, then does not contribute to the total visibility, but the flux fraction, $f_{\mathrm{FRC}}$, of that component still modifies $\epsilon$ via $\epsilon = f_{\mathrm{FDD}} / (f_{\mathrm{FDD}} + f_{\mathrm{FRC}})$. The radius $r$ in Fourier space is given by the spatial frequencies as $r = \sqrt{u^2+v^2}$. The spatial frequencies $u$ and $v$ are calculated from the projected baseline B, the position angle $\vartheta$, and the wavelength $\lambda$ via $u=B\sin{\vartheta}/\lambda$ and $v=B\cos{\vartheta}/\lambda$. The radial coordinate $\rho$ on the sky is defined by the angular coordinates $\alpha$ and $\beta$ (angular separation relative to the center of the object on the tangential sky plane) with $\rho = \sqrt{\alpha^2+\beta^2}$.

    The elliptical FDD, used in Sect.~\ref{secPhaseSubAsy} to analyze asymmetries, is derived from the circular FDD by applying a rotation (with an orientation given by the position angle $\vartheta$) and a compression to one of the axes (which becomes the minor axis). The rotation can be obtained by changing the Fourier and sky reference frames via
    \begin{eqnarray}
      u_{\vartheta} &=&  u \cos\vartheta - v \sin\vartheta \\[1mm]
      v_{\vartheta} &=&  u \sin\vartheta + v \cos\vartheta
    \end{eqnarray}
    \noindent and
    \begin{eqnarray}
      \alpha_{\vartheta} &=&  \alpha \cos\vartheta - \beta \sin\vartheta \\[1mm]
      \beta_{\vartheta}  &=&  \alpha \sin\vartheta + \beta \cos\vartheta \mathrm{,}
    \end{eqnarray}
    \noindent respectively.

    Applying the compression factor of $\eta = \theta_{\mathrm{FDD,min}} / \theta_{\mathrm{FDD,maj}}$, with $\theta_{\mathrm{FDD,min}}$ the minor diameter and $\theta_{\mathrm{FDD,maj}}$ the major diameter in Fourier space, yields the new variables
    \begin{eqnarray}
      r_{\vartheta,\eta}    &=& \sqrt{u_{\vartheta}^2\eta^2+v_{\vartheta}^2} \;\;\;\; \mathrm{and}\\
      \rho_{\vartheta,\eta} &=& \sqrt{\alpha_{\vartheta}^2/\eta^2+\beta_{\vartheta}^2} \mathrm{,}
    \end{eqnarray}
    \noindent respectively. The visibility and intensity functions then become
    \begin{eqnarray}
      V \; (r_{\vartheta,\eta}) &=& \left|\,\epsilon\;\;\frac{3\sqrt{\pi}\;J_{3/2}(\pi\theta_{\mathrm{FDD,maj}} r_{\vartheta,\eta})}{\sqrt{2}\;(\pi\theta_{\mathrm{FDD,maj}} r_{\vartheta,\eta})^{3/2}} \,\right|\;\mbox{and}\\[4mm]
      I \; (\rho_{\vartheta,\eta}) &=& \left\{ \begin{array}{ll}
        \epsilon\;\;\frac{8}{\eta\pi^2\theta_{\mathrm{FDD,maj}}^2}\; \sqrt{1-\left(\frac{2\rho_{\vartheta,\eta}}{\theta_{\mathrm{FDD,maj}}}\right)^2} & \mbox{if} \;\; \delta \leq 1\\[3mm]
         0 & \mbox{otherwise,} \end{array} \right.
    \end{eqnarray}
    \noindent respectively, with
    \begin{eqnarray}
      \delta &=& \frac{4 \alpha^2_{\vartheta}}{\theta^2_{\mathrm{FDD,min}}} + \frac{4 \beta^2_{\vartheta}}{\theta^2_{\mathrm{FDD,maj}}} \,\mbox{.}
    \end{eqnarray}

    Best-model parameters are derived by performing the Levenberg-Marquardt least-squares minimization procedure programmed for the interactive data language IDL as \texttt{MPFIT} by C.~B.~Markwardt\footnote{http://cow.physics.wisc.edu/$\sim$craigm/idl/idl.html}. To avoid the problem of running into local minima and to check for degeneracies, a global grid search was implemented. This is done in such a way that the least-squares minimization always operates only between two grid points in each parameter, while going through the whole parameter space. Formal 1$\,\sigma$ errors were computed by taking the difference between the best fit value and the value given at the lowest non-reduced chi-square plus one, knowing that this is only an approximation, since the chi-square values are not Gaussian distributed. To get a more appropriate parameter error, the visibility errors are scaled to a value yielding a reduced chi-square of one. Besides this, the final parameter error is always a multiple of the shortest grid distance. This approach leads to an adequate description of the data by an FDD and to reasonable error estimates.

%%%%%%%%%%%%%%%%%%%%%%%%%%%%%%%%%%%%%%%%%%%%%%%%%%%%%%%%%%%%%%%%%%%%%%%%%%%%%%%%%%%%%%%%%%%%
\subsection{Model fitting results}\label{secModSubModel}

    The left hand panel of Fig.~\ref{FigSymModel} shows the fit of a circular FDD and UD to all 75~visibility measurements for wavelength bin~6 (9.07~$\mu$m). Clearly, the second lobe is better fitted by an FDD. A comparison of the reduced chi-square values verified that a UD is not a good representation of the brightness distribution of W~Hya. The best-fit parameters of the circular FDD are given in Table~\ref{TableResults} for all 25~wavelength bins. Formal errors are computed as mentioned in the previous section, while the listed mean visibility errors are derived as described in Sect.~\ref{secObsSubRed}. A residual analysis shows that for each wavelength bin on average about 60\% of the measurements are inside the 3$\,\sigma$ range. This low value stems from using rather low uncertainties for the visibilities and this fit not accounting for the scatter of the data due to the influences of the pulsation phase and the asymmetry.

%---------------------------------------------------------------
   \begin{table*}
     \caption[]{Results of the circular and elliptical fully limb-darkened disk fits.}
     \label{TableResults}
     \centering
     \begin{tabular}{cc|cc|cccc}
       \hline
       \noalign{\smallskip}
       & & \multicolumn{2}{c}{circular FDD}           &   \multicolumn{4}{c}{elliptical FDD} \\
       Wavelength              & Visibility           & $\theta_{\mathrm{FDD}}$ & Flux $\epsilon$   & 
       $\theta_{\mathrm{FDD,maj}}$ & Axis Ratio $\eta$    & PA $\vartheta$   & Flux $\epsilon$   \\
       ($\mu\mathrm{m}$)       & error$^{\mathrm{a}}$ & ($\mathrm{mas}$) &                   & 
       ($\mathrm{mas}$)        &                      & ($^\circ$)       &                   \\
       \noalign{\smallskip}
       \hline
       \noalign{\smallskip}
$\;$ 8.12& 0.007& $\;$ 81.0 $\pm$ 1.0& 0.86 $\pm$ 0.02& $\;$ 92.5 $\pm$ 5.0& 0.87 $\pm$ 0.07 &  $-1$  $\pm$ 4 & 0.89 $\pm$ 0.03 \\
$\;$ 8.33& 0.007& $\;$ 79.9 $\pm$ 1.2& 0.84 $\pm$ 0.01& $\;$ 88.4 $\pm$ 5.0& 0.89 $\pm$ 0.08 &$\;\,$6 $\pm$ 4 & 0.87 $\pm$ 0.03 \\
$\;$ 8.54& 0.006& $\;$ 80.8 $\pm$ 1.0& 0.83 $\pm$ 0.01& $\;$ 89.3 $\pm$ 5.0& 0.89 $\pm$ 0.07 &  $-1$  $\pm$ 4 & 0.84 $\pm$ 0.03 \\
$\;$ 8.71& 0.006& $\;$ 80.9 $\pm$ 1.6& 0.86 $\pm$ 0.02& $\;$ 90.5 $\pm$ 5.0& 0.86 $\pm$ 0.07 &  $-1$  $\pm$ 4 & 0.86 $\pm$ 0.03 \\
$\;$ 8.87& 0.006& $\;$ 79.6 $\pm$ 1.0& 0.86 $\pm$ 0.01& $\;$ 84.9 $\pm$ 5.0& 0.91 $\pm$ 0.08 &    14  $\pm$ 4 & 0.86 $\pm$ 0.03 \\
$\;$ 9.07& 0.006& $\;$ 79.2 $\pm$ 1.0& 0.85 $\pm$ 0.01& $\;$ 85.3 $\pm$ 5.0& 0.89 $\pm$ 0.08 &    15  $\pm$ 4 & 0.86 $\pm$ 0.03 \\
$\;$ 9.26& 0.006& $\;$ 78.0 $\pm$ 1.0& 0.83 $\pm$ 0.02& $\;$ 85.1 $\pm$ 5.0& 0.88 $\pm$ 0.08 &    18  $\pm$ 4 & 0.85 $\pm$ 0.03 \\
$\;$ 9.45& 0.006& $\;$ 77.9 $\pm$ 1.2& 0.81 $\pm$ 0.02& $\;$ 87.5 $\pm$ 5.0& 0.86 $\pm$ 0.08 &    13  $\pm$ 4 & 0.83 $\pm$ 0.03 \\
$\;$ 9.63& 0.006& $\;$ 79.6 $\pm$ 1.2& 0.80 $\pm$ 0.02& $\;$ 89.4 $\pm$ 5.0& 0.85 $\pm$ 0.07 &    14  $\pm$ 4 & 0.81 $\pm$ 0.03 \\
$\;$ 9.78& 0.006& $\;$ 79.8 $\pm$ 1.2& 0.79 $\pm$ 0.02& $\;$ 90.5 $\pm$ 5.0& 0.84 $\pm$ 0.07 &    14  $\pm$ 4 & 0.81 $\pm$ 0.03 \\
$\;$ 9.92& 0.006& $\;$ 79.0 $\pm$ 1.2& 0.78 $\pm$ 0.01& $\;$ 89.7 $\pm$ 5.0& 0.85 $\pm$ 0.07 &$\;\,$3 $\pm$ 4 & 0.78 $\pm$ 0.03 \\
    10.09& 0.005& $\;$ 78.8 $\pm$ 1.0& 0.77 $\pm$ 0.01& $\;$ 89.7 $\pm$ 5.0& 0.85 $\pm$ 0.07 &$\;\,$3 $\pm$ 4 & 0.78 $\pm$ 0.03 \\
    10.26& 0.005& $\;$ 80.4 $\pm$ 1.0& 0.76 $\pm$ 0.01& $\;$ 91.5 $\pm$ 5.0& 0.85 $\pm$ 0.07 &$\;\,$5 $\pm$ 4 & 0.77 $\pm$ 0.03 \\
    10.42& 0.005& $\;$ 84.5 $\pm$ 1.4& 0.77 $\pm$ 0.01& $\;$ 93.7 $\pm$ 5.0& 0.86 $\pm$ 0.07 &    13  $\pm$ 4 & 0.77 $\pm$ 0.03 \\
    10.58& 0.005& $\;$ 87.3 $\pm$ 1.4& 0.77 $\pm$ 0.02& $\;$ 98.1 $\pm$ 5.0& 0.84 $\pm$ 0.07 &    15  $\pm$ 4 & 0.78 $\pm$ 0.03 \\
    10.71& 0.005& $\;$ 89.2 $\pm$ 1.4& 0.77 $\pm$ 0.01& $\;$ 99.6 $\pm$ 5.0& 0.85 $\pm$ 0.07 &    15  $\pm$ 4 & 0.78 $\pm$ 0.03 \\
    10.84& 0.005& $\;$ 91.2 $\pm$ 1.2& 0.76 $\pm$ 0.01&     102.2 $\pm$ 5.0& 0.86 $\pm$ 0.06 &    11  $\pm$ 4 & 0.77 $\pm$ 0.03 \\
    10.99& 0.004& $\;$ 94.8 $\pm$ 1.4& 0.76 $\pm$ 0.01&     104.5 $\pm$ 5.0& 0.86 $\pm$ 0.06 &    15  $\pm$ 4 & 0.76 $\pm$ 0.03 \\
    11.14& 0.004& $\;$ 96.3 $\pm$ 1.2& 0.76 $\pm$ 0.01&     105.7 $\pm$ 5.0& 0.87 $\pm$ 0.06 &    14  $\pm$ 4 & 0.77 $\pm$ 0.03 \\
    11.29& 0.004& $\;$ 97.8 $\pm$ 1.2& 0.77 $\pm$ 0.01&     106.8 $\pm$ 5.0& 0.88 $\pm$ 0.06 &    15  $\pm$ 4 & 0.77 $\pm$ 0.03 \\
    11.43& 0.005& $\;$ 98.9 $\pm$ 1.4& 0.76 $\pm$ 0.01&     108.2 $\pm$ 5.0& 0.87 $\pm$ 0.06 &    15  $\pm$ 4 & 0.77 $\pm$ 0.03 \\
    11.55& 0.005&     100.5 $\pm$ 1.4& 0.77 $\pm$ 0.01&     109.8 $\pm$ 5.0& 0.88 $\pm$ 0.06 &    14  $\pm$ 4 & 0.78 $\pm$ 0.03 \\
    11.67& 0.005&     101.6 $\pm$ 1.4& 0.77 $\pm$ 0.01&     110.8 $\pm$ 5.0& 0.88 $\pm$ 0.06 &    15  $\pm$ 4 & 0.78 $\pm$ 0.03 \\
    11.81& 0.004&     103.4 $\pm$ 1.4& 0.78 $\pm$ 0.02&     112.3 $\pm$ 5.0& 0.88 $\pm$ 0.06 &    16  $\pm$ 4 & 0.79 $\pm$ 0.03 \\
    11.95& 0.004&     104.7 $\pm$ 1.2& 0.79 $\pm$ 0.01&     113.2 $\pm$ 5.0& 0.88 $\pm$ 0.06 &    19  $\pm$ 4 & 0.80 $\pm$ 0.03 \\
       \noalign{\smallskip}
       \hline
      \end{tabular}
      \begin{list}{}{}
      \item[$^{\mathrm{a}}$] This is the mean visibility error used for the corresponding wavelength bin. See Sect.~\ref{secObsSubRed} for the definition of this value.
      \end{list}
    \end{table*}
%---------------------------------------------------------------

\begin{figure*}
     \centering
     \includegraphics[width=0.49\linewidth]{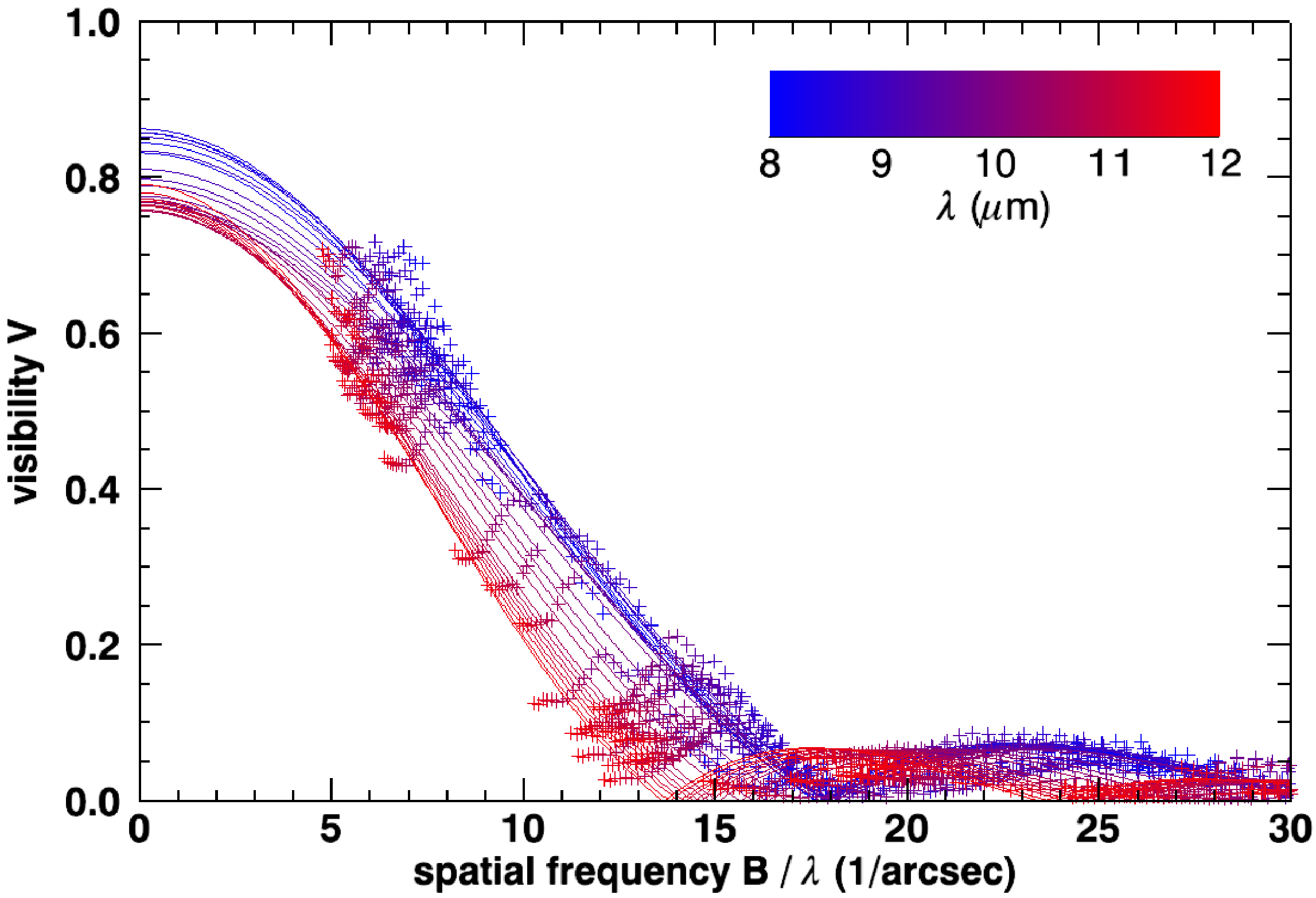}
     \includegraphics[width=0.49\linewidth]{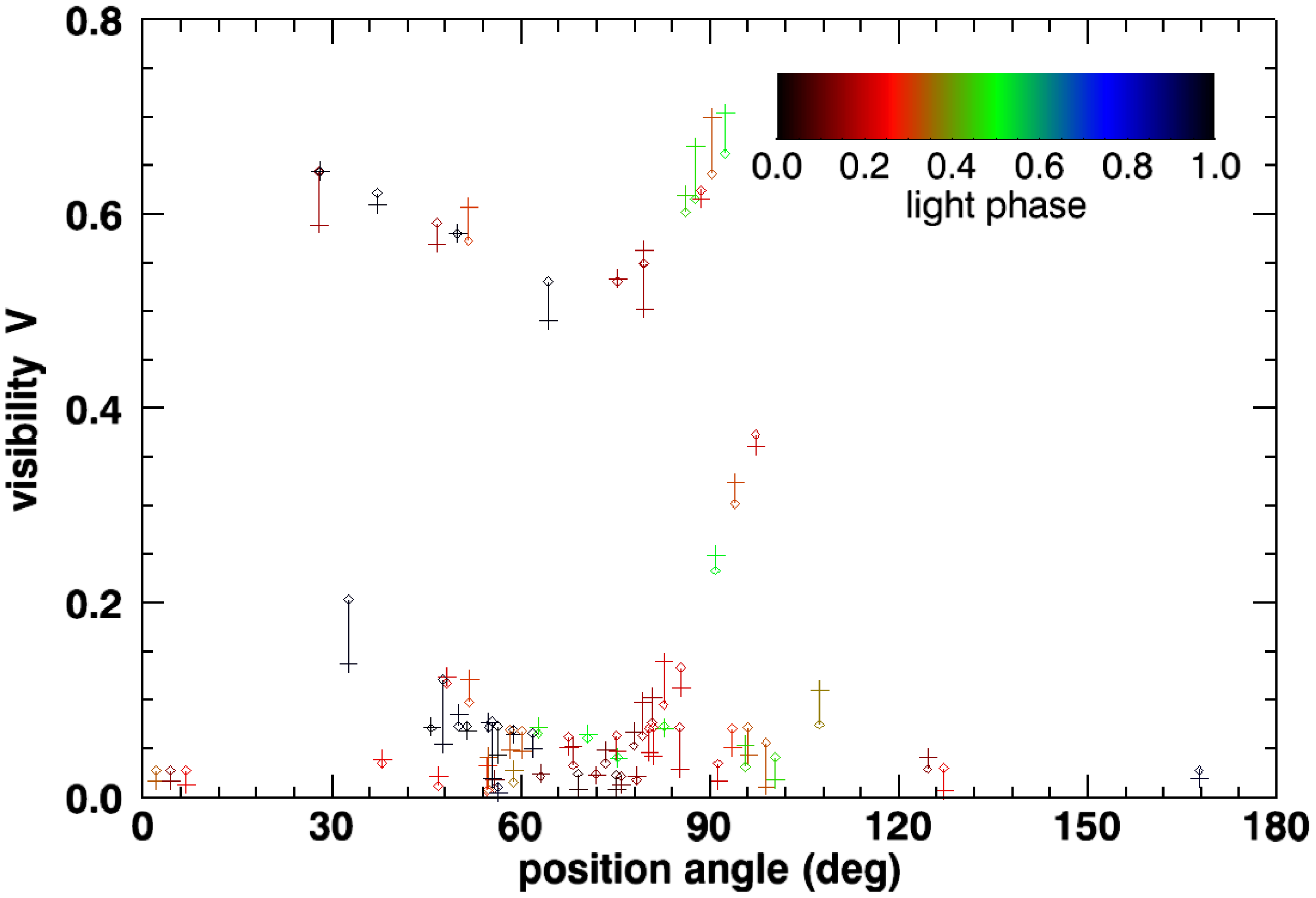}
     \caption{\textit{Left:} Same as left hand panel of Fig.~\ref{FigSymModel}, but for all 25~wavelength bins. Clearly the first zero shifts to lower spatial frequencies with increasing wavelength, i.e.~the FDD diameter increases with increasing wavelength. \textit{Right:} Visibility as function of position angle for wavelength bin~6 (9.07~$\mu$m). Measurements (crosses) are connected with their corresponding circular FDD model points (small open circles) showing that there is a systematic shift in the difference, $V_{\mathrm{model}} - V_{\mathrm{measurement}}$, by going from 0$^\circ$ to 90$^\circ$. For data points with low visibilities, corresponding to data in the increasing part of the second lobe, the systematic is inverse. See Sect.~\ref{secPhase} for a detailed interpretation.}
     \label{Fig3D}
   \end{figure*}

\begin{figure*}
     \centering
     \includegraphics[width=0.49\linewidth]{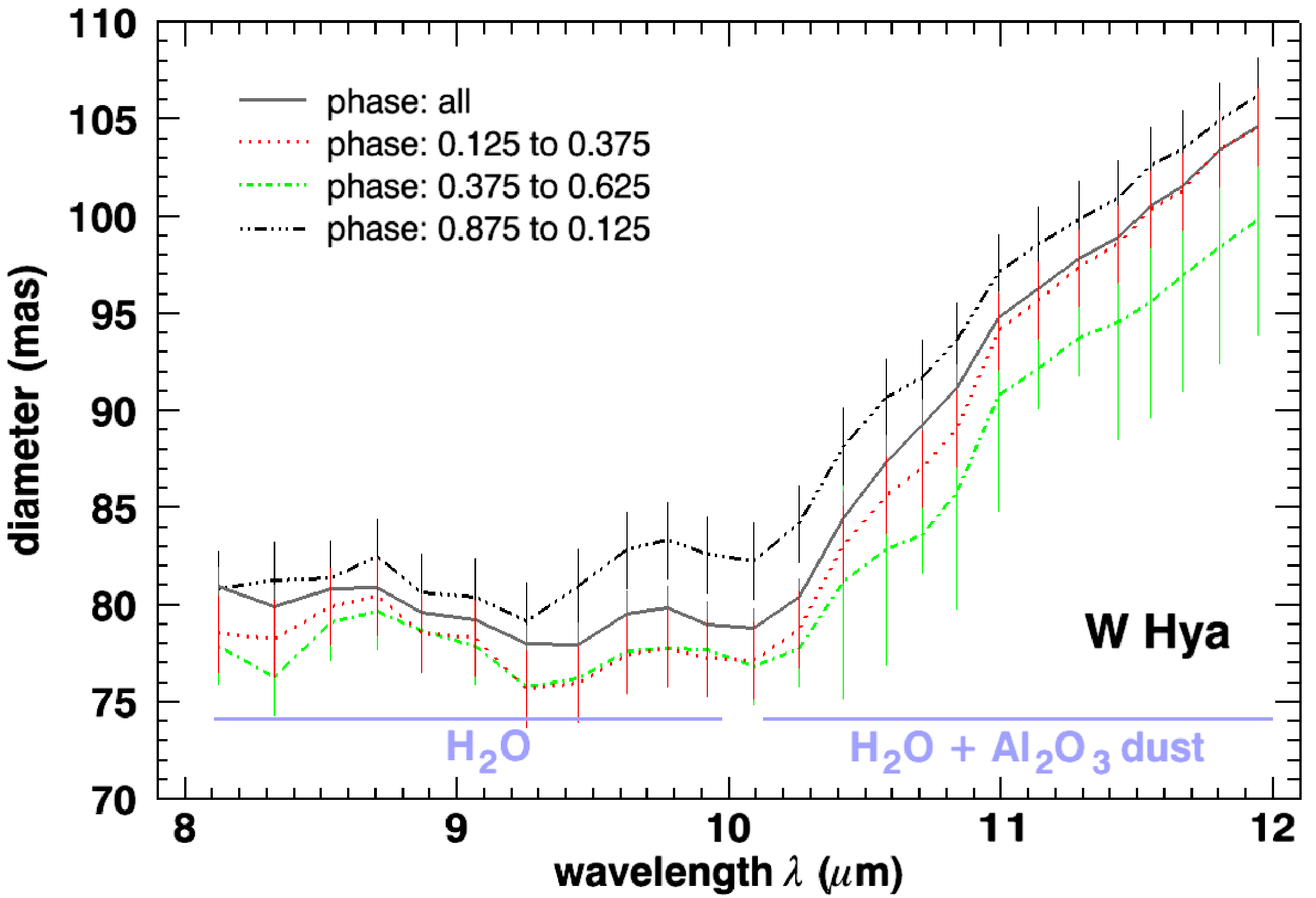}
     \includegraphics[width=0.49\linewidth]{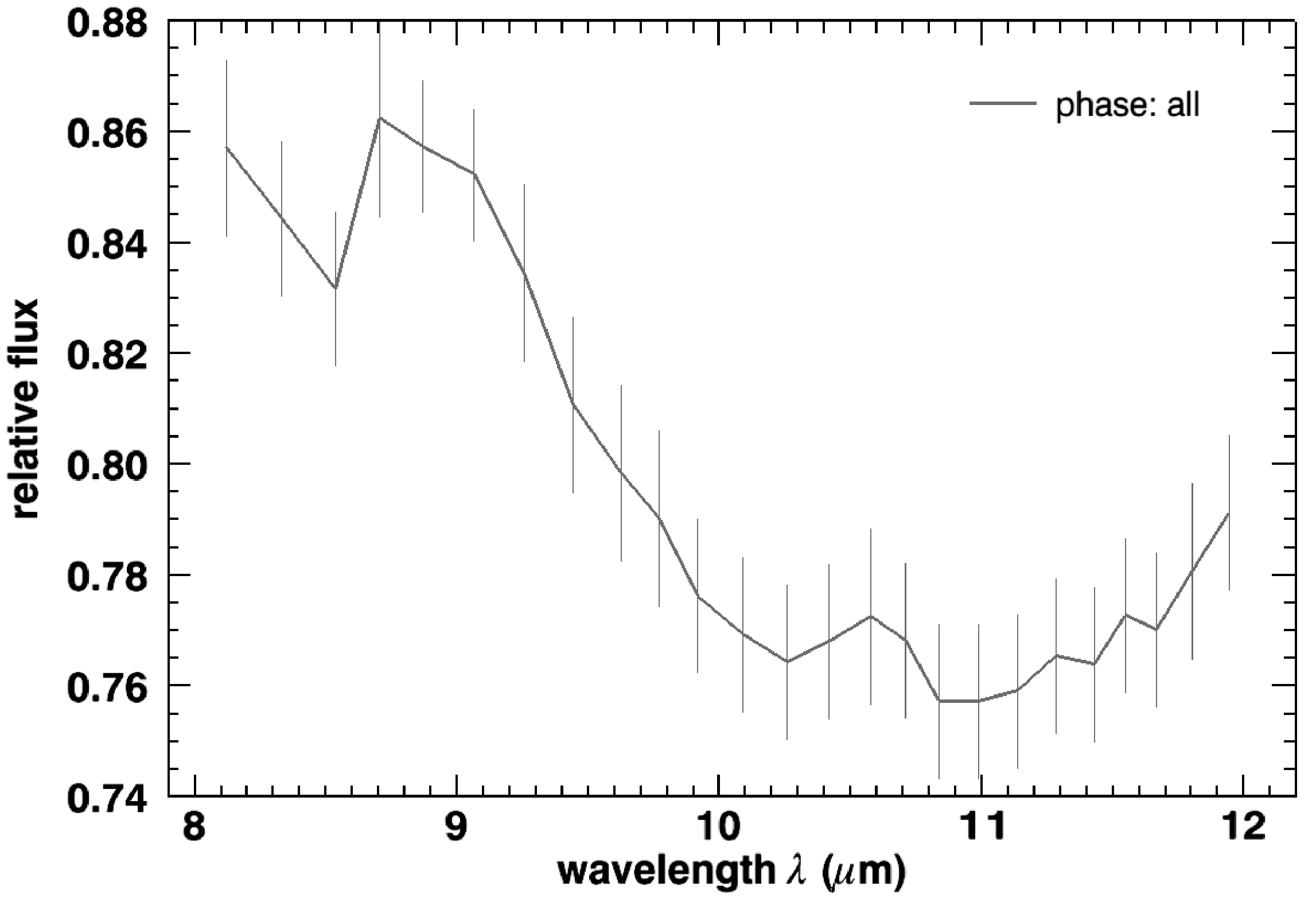}
     \caption{\textit{Left:} The fitted circular fully limb-darkened disk diameter, $\theta_{\mathrm{FDD}}$, as function of wavelength. The solid gray line shows the fit to all 75~measurements and the other lines the fits to individual pulsation phase bins. \textit{Right:} The relative flux contribution,~$\epsilon$, of the circular FDD obtained for the fit to the full data set. These values are fixed for the fit of the circular FDD to the individual pulsation phase bins.}
     \label{FigPara}
   \end{figure*}

    The left hand panel of Fig.~\ref{Fig3D} shows the fit to all 25 wavelength bins. Clearly, the first zero shifts to lower spatial frequencies with increasing wavelength, i.e.~the FDD diameter,~$\theta_{\mathrm{FDD}}$, increases with increasing wavelength. At the same time, the relative flux contribution, $\epsilon$, of the FDD decreases. This behavior is shown more exactly in the two parameter plots in Fig.~\ref{FigPara}. From these panels it can be derived that $\theta_{\mathrm{FDD}}$ actually stays constant at a value of about (80~$\pm$~1.2)~mas between 8 and 10~$\mu$m, while it gradually increases at wavelengths longer than 10~$\mu$m to reach (105~$\pm$~1.2)~mas at 12~$\mu$m. The diameter increase in the longer wavelength regime corresponds to a relative increase, $\theta_{FDD,12\mu\mathrm{m}}$/$\theta_{FDD,10\mu\mathrm{m}}$, of (31~$\pm$~3)\%. This apparent increase from 80~mas to 105~mas is equivalent to an increase from 7.1~AU to 9.5~AU at the distance of W~Hya. The molecules and close dust species causing this shape are already indicated inside the plot, but will be discussed further in Sect.~\ref{secModSubRes}. In contrast, the relative flux decreases from (0.85~$\pm$~0.02) to (0.77~$\pm$~0.02), reflecting the increased flux contribution from the fully resolved colder surrounding silicate dust shell going to longer wavelengths. A fit to the unbinned grism data results in the same shapes and does not reveal any additional features.

%---------------------------------------------------------------
%---------------------------------------------------------------
%---------------------------------------------------------------
%---------------------------------------------------------------

%%%%%%%%%%%%%%%%%%%%%%%%%%%%%%%%%%%%%%%%%%%%%%%%%%%%%%%%%%%%%%%%%%%%%%%%%%%%%%%%%%%%%%%%%%%%
\subsection{Wavelength dependence of the diameter}\label{secPhaseSubWave}

\begin{figure*}
     \centering
     \includegraphics[width=0.9\linewidth]{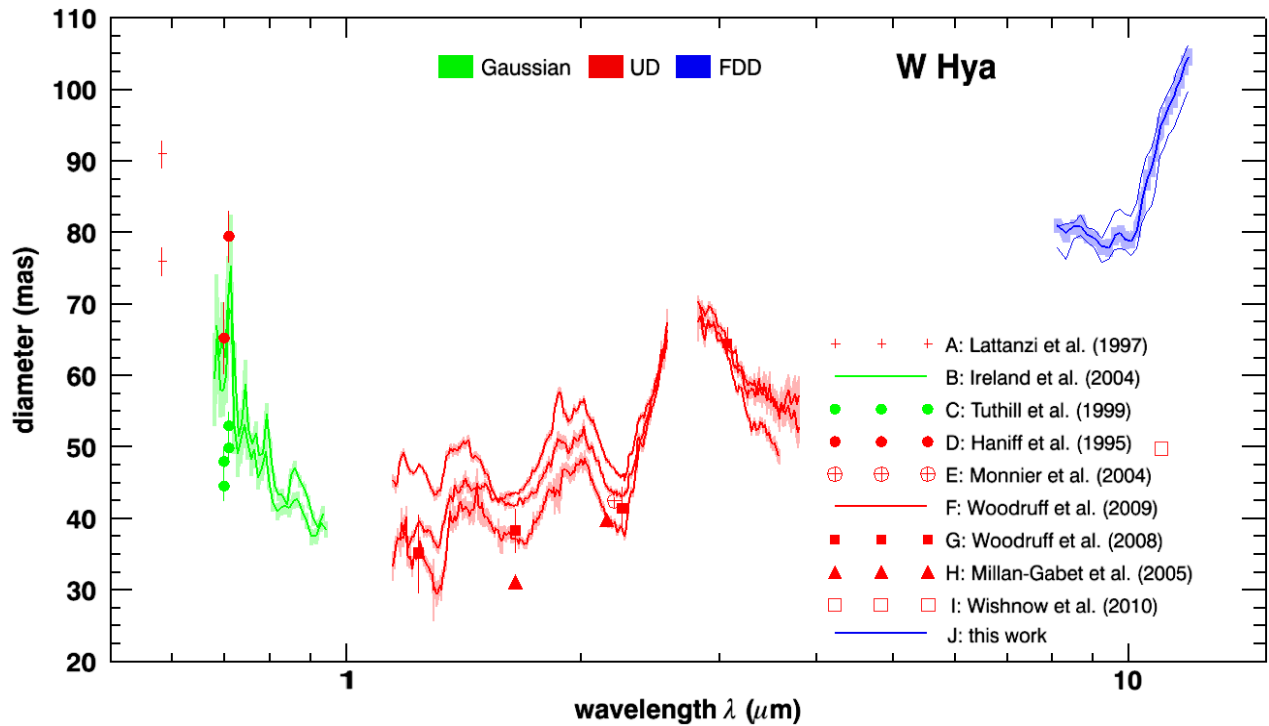}
     \caption{Diameter measurements of W~Hya over a wide wavelength range with different models reported by various authors. In addition to the N-band full data set diameter (thick blue line; all 75~observations), the diameters at maximum light (upper thin blue line) and minimum light (lower thin blue line) from this work are shown as well (FDD $\approx$ $1.15$~UD $\approx$ $1.68$~Gaussian). See Table~\ref{TableAuthor} for more details.}
     \label{FigDia}
   \end{figure*}

    Before interpreting the previously obtained FDD diameter behavior in the next section, it is set in relation to measurements at other wavelengths. Figure~\ref{FigDia} shows interferometric the diameter determinations reported by various authors from the visual to the mid-IR. They were obtained by fitting a Gaussian (green), a uniform disk (UD, red) or a fully limb-darkened disk (FDD, blue). A conversion between the models is not performed, since the various diameter determinations depend on the spatial frequency coverage\footnote{Empirical conversion factors are approximately: FDD $\approx$ 1.15~UD $\approx$ 1.68~$FWHM$ (assuming a uniform spatial frequency coverage in the first lobe).}. Information on visual light phases and position angles (if applicable) of these observations can be found in Table~\ref{TableAuthor}. In addition to the FDD diameter, derived in this study from the fit to the full data set, the diameter trends at visual maximum (upper curve) and visual minimum (lower curve) are plotted as well (cf.~Sect.~\ref{secPhaseSubLight}).
    
    The observed apparent diameter changes dramatically within the given wavelength range due to the strong wavelength-dependent opacities of the atmospheric constituents (cf.~e.g.~Baschek et al.~\cite{Baschek_et91} and Scholz et al.~\cite{Scholz01}). In the optical, the measured diameters are sensitive to TiO bands. The largest variations are around the strongest bands at 712~nm and 670~nm with apparent diameter enlargements of up to a factor of~2. It is notable that the diameters observed at around visual minimum by Ireland et al.~(\cite{Ireland04}) are systematically larger compared to the diameters obtained at around visual maximum by Tuthill et al.~(\cite{Tuthill99}). Ireland et al.~(\cite{Ireland04}) conclude that significant dust production occurs near minimum, and suggest that the large increase in apparent diameter towards the blue is caused by light scattered by dust. Gaussian diameters between 0.7 and 1.0~$\mu$m were measured by Ireland et al.~(\cite{Ireland04}) at two position angles ($120^{\circ}$ and $252^{\circ}$). Even though the diameters at both position angles are slightly different, the authors conclude that this is not significant.

    In the near-IR, H$_2$O and CO in different layers are predominantly responsible for the wavelength dependence of the diameter. Most of the observations are conducted at J, H, and K~bands (1.25, 1.65, and 2.2~$\mu$m, respectively). It can be inferred that also in the near-IR diameters vary by a factor of~2 and that there is a complex diameter dependence on pulsation cycle and pulsation phase (cf.~ e.g.~Weiner et al.~\cite{Weiner_et03}; Woodruff et al.~\cite{Woodruff09}). The gap at around 2.7~$\mu$m in the curve of Woodruff et al.~(\cite{Woodruff09}) comes from telluric contamination. A reasonable estimate for the true photospheric diameter of an AGB star can be obtained from uniform disk measurements at K-Band. The 2.2~$\mu$m diameter is approximately 1.2~times the true photospheric diameter (e.g.~Millan-Gabet et al.~\cite{Millan05}). 
    
    Wishnow et al.~(\cite{Wishnow_et10}) assume a UD diameter of the stellar component in the mid-IR (11.15~$\mu$m) of about 50~mas for their model. However, this value was reported as not very accurate. Since the FDD diameters obtained with MIDI describe a region whose exact location depends on the flux contribution of all constituents (continuum photosphere, atmospheric molecular layers, and nearby dust shells; cf.~Sect.~\ref{secModSubRes}) as function of wavelength and pulsation phase, and not only the photosphere of the star, both diameters are not comparable, and the apparent discrepancy can be explained. The average of all diameters at 2.2~$\mu$m of the authors given in Table~\ref{TableAuthor} is (42.8~$\pm$~3.5)~mas. This gives a ratio, $\theta_{\mathrm{FDD},10\mu\mathrm{m}}$/$\theta_{\mathrm{UD},2.2\mu\mathrm{m}}$, of 1.8~$\pm$~0.2. The UD fit value should be used to compare similar models, and the ratio, $\theta_{\mathrm{UD},10\mu\mathrm{m}}$/$\theta_{\mathrm{UD},2.2\mu\mathrm{m}}$, becomes 1.6~$\pm$~0.2. From the mean K-band diameter the photospheric diameter, $\theta_{\mathrm{phot}}$, can be estimated to about 36~mas (42.8~mas divided by 1.2, cf.~previous paragraph). This value is assumed for $\theta_{\mathrm{phot}}$ throughout this paper and gives a mid-IR to photospheric diameter ratio, $\theta_{\mathrm{UD},10\mu\mathrm{m}}$/$\theta_{\mathrm{phot}}$, of 1.9~$\pm$~0.2.

    Maser observations of different molecules give additional diameter constraints. SiO and H$_2$O masers probe inner regions where the molecular spheres are present and dust formation takes place, while OH masers trace wind regions farther out. Ring diameters for SiO masers were determined by Miyoshi et al.~(\cite{Miyoshi_et94}), Cotton  et al.~(\cite{Cotton_et04}, \cite{Cotton_et08}), Reid~\& Menten~(\cite{Reid_Menten07}), and Imai et al.~(\cite{Imai_et10}). The average of their diameters is about (77~$\pm$~11)~mas, which gives a ratio, $\theta_{\mathrm{SiO}}$/$\theta_{\mathrm{FDD},10\mu\mathrm{m}}$, of 1.0~$\pm$~0.1 ($\theta_{\mathrm{SiO}}$/$\theta_{\mathrm{UD},10\mu\mathrm{m}}$~= 0.9~$\pm$~0.1). H$_2$O maser ring diameters for W~Hya are on the order of 300~mas (Reid \& Menten \cite{Reid_Menten90}), while OH maser ring diameters at 1665 and 1667~MHz were determined to about 700 and 1130~mas, respectively (Szymczak et al.~\cite{Szymczak_et98}, but see also Shintani et al.~\cite{Shintani_et08}). In addition, thermal line emissions of HCN (Muller et al.~\cite{Muller_et08} and Ziurys et al.~\cite{Ziurys_et09}) and CO (e.g.~Olofsson et al.~\cite{Olofsson_et02} and Ziurys et al.~\cite{Ziurys_et09}) were repeatedly detected as well.

%---------------------------------------------------------------
%---------------------------------------------------------------
%---------------------------------------------------------------
   \begin{table*}
     \caption{Details to the diameter measurements presented in Fig.~\ref{FigDia}.}
     \label{TableAuthor}
     \centering
     \begin{tabular}{llccl}
       \hline
       \noalign{\smallskip}
       & Author & Model & Phase$^{\mathrm{a}}$ & Comments and position angles (PA) \\
       \noalign{\smallskip}
       \hline
       \noalign{\smallskip}
         A:& Lattanzi et al.~(\cite{Lattanzi97})  & UD       & 0.64 & two perpendicular axes ($\eta$~$\approx$~0.86, PA~$\approx$~$143^{\circ}$, 583~nm)\\
         B:& Ireland et al.~(\cite{Ireland04})    & Gaussian & 0.44 & upper curve for PA~$120^{\circ}$ and lower curve for PA~$252^{\circ}$ ($0.7-1.0$~$\mu$m)\\
         C:& Haniff et al.~(\cite{Haniff_et95})   & UD       & 0.04 & uniform disk diameter at 700 and 710~nm\\
         D:& Tuthill et al.~(\cite{Tuthill99})    & Gaussian & 0.04 & elliptical Gaussian at 700 and 710~nm ($\eta$~$\approx$~0.94, PA~$\approx$~$93^{\circ}$)\\
         E:& Monnier et al.~(\cite{Monnier04})    & UD       & 0.50 & UD model was reported as a bad fit to the data (K band)\\
         F:& Woodruff et al.~(\cite{Woodruff09})  & UD  &  $0.58-1.53$ & curves for phase 0.58~(middle), 0.79~(lower) and 1.53~(upper)\\
         G:& Woodruff et al.~(\cite{Woodruff08})  & UD  &  $0.50-1.00$ & mean of multiple measurements in given phase range (J, H, K, \& L band)\\
         H:& Millan-Gabet et al.~(\cite{Millan05})& UD  &      0.60    & UD diameter for H and K bands\\
         I:& Wishnow et al.~(\cite{Wishnow_et10}) & UD  &   $0.1-1.1$  & assumed photospheric UD diameter at 11.15~$\mu$m, reported as not very accurate\\
         J:& this work                            & FDD & 0.00 \& 0.50 & curves for full data set (middle), max (upper), and min (lower) light\\
       \noalign{\smallskip}
       \hline
     \end{tabular}
    \begin{flushleft}
       $^{\mathrm{a}}$~visual light phase (phases refer to the phases determined by the original authors)
     \end{flushleft}
   \end{table*}
%---------------------------------------------------------------

%%%%%%%%%%%%%%%%%%%%%%%%%%%%%%%%%%%%%%%%%%%%%%%%%%%%%%%%%%%%%%%%%%%%%%%%%%%%%%%%%%%%%%%%%%%%
\subsection{Discussion and interpretation}\label{secModSubRes}

    Studying the interplay of the extended pulsating atmosphere with molecular spheres directly above and the dust formation and wind acceleration zone farther out is crucial for understanding the physical and chemical processes working in AGB stars. With the MIDI interferometer, the photosphere and molecular layers of W~Hya are probed, sampling also the region where the first seeds for dust formation originate.

    The spectrum in Fig.~\ref{FigSpec} shows that molecules, such as CO, H$_2$O, and SiO, are present in the upper atmosphere. In the N-band between 8 and 13~$\mu$m, strong pure-rotation lines of H$_2$O are expected. In addition, SiO exhibits fundamental bands between 8 and 10~$\mu$m (Decin~\cite{Decin00}). Modeling by Justtanont et al.~(\cite{Justtanont_et04}) also reveals at least three different dust species, namely amorphous silicate, Al$_2$O$_3$, and MgFeO, which are likely located at different distances from the star. In particular, amorphous Al$_2$O$_3$ provides significant opacity for wavelengths longwards of 10~$\mu$m (cf.~e.g.~Koike et al.~\cite{Koike_et95}, Begemann et al.~\cite{Begemann_et97}, Posch et al.~\cite{Posch_et99}, Egan \& Sloan~\cite{Egan_Sloan01}, Woitke et al.~\cite{Woitke06}, Ireland \& Scholz~\cite{Ireland_Scholz06} and Robinson \& Maldoni~\cite{Robinson_Maldoni10}), and it can survive at high temperatures.

    Quasi-static, warm, and dense molecular layers close to the star, at typically $2-3$~photospheric radii ($R_{\mathrm{phot}}$), are detected for O-rich (e.g.~Mennesson et al.~\cite{Mennesson_et02}; Perrin et al.~\cite{Perrin_et04}; Ireland et al.~\cite{Ireland_et04d}; Woodruff et al.~\cite{Woodruff_et04} and Fedele et al.~\cite{Fedele_et2005}) and C-rich (e.g.~Hron et al.~\cite{Hron_et98} and Ohnaka et al.~\cite{Ohnaka_et07}) AGB stars, as well as for RSG stars (e.g.~Perrin et al.~\cite{Perrin_et07}). These layers were introduced earlier to explain spectroscopic observations (e.g.~Hinkle \& Barnes~\cite{Hinkle_Barnes_79}; Tsuji et al.~\cite{Tsuji_et97} and Yamamura et al.~\cite{Yamamura_et99}).
    
    In particular, the O-rich Miras RR~Sco (Ohnaka et al.~\cite{Ohnaka_et05}) and S~Ori (Wittkowski et al.~\cite{Wittkowski_et07}) show a diameter behavior very similar to that of W~Hya throughout the N-band. Comparing the shape from the left hand panel of Fig.~\ref{FigPara} with Fig.~1e in Ohnaka et al.~(\cite{Ohnaka_et05}) and Figs.~$2-5$d in Wittkowski et al.~(\cite{Wittkowski_et07}) leads to the idea that, in all three stars, the same constituents and mechanisms are responsible for this appearance.
    
    The diameter ratio $\theta_{12\mu\mathrm{m}}$/$\theta_{10\mu\mathrm{m}}$ is approximately~1.3 for RR~Sco, for S~Ori between 1.3 and 1.5, and for W~Hya approximately~1.3 (Sect.~\ref{secModSubModel}). Even the increase between the K-band diameter and the diameter at 10~$\mu$m is similar. The ratio $\theta_{\mathrm{UD},10 \mu\mathrm{m}}$/$\theta_{\mathrm{UD},2.2 \mu\mathrm{m}}$ is for RR~Sco approximately 1.8, for S~Ori between 1.6 and 2.2 and for W~Hya approximately~1.6 (Sect.~\ref{secPhaseSubWave}).

    Ohnaka et al.~(\cite{Ohnaka_et05}) modeled RR~Sco by adding a molecular layer of H$_2$O and SiO, and a dust layer, consisting of aluminum oxide and silicates, to a static star with a fixed temperature. This composition reproduced the N-band diameter behavior very reasonably, with an inner radius for the dust shell of 7.5~$R_{\mathrm{phot}}$, while the molecular layers were located at a radius of 2.3~$R_{\mathrm{phot}}$. The large dust shell radius might be due to the mix of 80\% aluminum oxide (Al$_2$O$_3$) and 20\% silicates and because the same density profile and inner condensation radius were used for both dust species.

    Wittkowski et al.~(\cite{Wittkowski_et07}) modeled S~Ori with the dynamic atmospheric models of Ireland et al.~(\cite{Ireland_et04b}, \cite{Ireland_et04c}), where the molecular layers are naturally included and only the dust shell has to be added ad hoc. This assumes that the stellar atmosphere and dust shell are spatially separated. They allowed different density profiles and condensation radii for the aluminum oxide and silicate dust. The model without silicates fitted the data better. The inner boundary of the Al$_2$O$_3$ dust shell varied between 1.8 and 2.4 photospheric radii. At visual minimum, the dust shell was more compact with a larger optical depth, and it had a smaller radius, while at visual maximum the opposite was the case.
    
    W~Hya can be explained by a similar model. The overall larger diameter in the mid-IR is caused by a warm molecular layer of H$_2$O, and the gradual increase longward of 10~$\mu$m arises most likely from the presence of Al$_2$O$_3$ dust close to this layer. A nearby SiO molecular shell might be of some relevance for the diameter enlargement between 8 and 10~$\mu$m as well. The formation of Al$_2$O$_3$ dust at these short distances from the stellar surface would be consistent with the empirical results by Lorenz-Martins \& Pompeia (\cite{LorenzMartins_Pompeia00}), as well as with recent theoretical calculations by Ireland \& Scholz (\cite{Ireland_Scholz06}) and Woitke (\cite{Woitke06}).
    
    Dust can only exist close to the star if the temperature is below the condensation temperature of the dust species. A rough estimate of the equivalent blackbody temperature from the total MIDI flux, the flux fraction~$\epsilon$, and the diameter~$\theta_{\mathrm{FDD}}$ gives a value of about (800~$\pm$~100)~K at the location of the FDD radius. This temperature is low enough that aluminum oxide dust can condensate (cf.~e.g.~Woitke~\cite{Woitke_et99}). Since Al$_2$O$_3$ has only a moderate abundance and a low absorption efficiency at optical and near-IR wavelengths, i.e.~near the flux maximum at around 1~$\mu$m, it cannot be responsible for initiating the mass loss (cf.~e.g.~Woitke~\cite{Woitke06}). Al$_2$O$_3$ can exist close to the star without inducing mass loss.

    Scattering off large Fe-free silicate grains may solve this problem, as these micron-sized silicates do not have a considerable absorption cross-section, but a high radiative scattering cross-section around the flux maximum at 1~$\mu$m (cf.~H{\"o}fner~\cite{Hoefner08}). However, these grains are not detected with MIDI. If a large amount of Fe-free silicates existed \textit{close} to the star, an emission feature in the MIDI spectrum at around 10~$\mu$m would be present (due to strong vibrational resonances), and it would also modify the apparent FDD diameter around this wavelength. Therefore, the wind acceleration mechanism proposed by H{\"o}fner~(\cite{Hoefner08}) cannot be directly verified. However, since the mass-loss rate of W~Hya is very low (cf.~e.g.~Nowotny et al.~\cite{Nowotny_et10}), this behavior is not surprising. % and silicates are not a major constituent of the close dust layer

    This applies not only to W~Hya, but also to RR~Sco and S~Ori. All three O-rich AGB stars have moderate to low mass-loss rates. The similar behavior of the apparent diameter throughout the N-band suggests that the appropriate mechanism for a moderate-to-low mass-loss rate is similar, since the conditions appear similar in the transition region (between the photosphere and the silicate dust shell) where the wind should be initiated, even if the mechanism cannot be clarified here.%This favors the model proposed by H{\"o}fner \& Anderson~(\cite{HoefnerAndersen_07}).

    In general, another possibility of forming a wind in O-rich AGB stars is to accelerate small amounts of carbon grains (H{\"o}fner \& Anderson~\cite{HoefnerAndersen_07}). Unfortunately, these carbon grains, formed in nonequilibrium environments, do not show any spectral features in the mid-infrared and would probably produce IR colors that are not consistent with the observations (H{\"o}fner, private communication). In this context it might be interesting to know whether scattering on Al$_2$O$_3$ grains is important, even if they are probably not abundant enough. Also the role of large amounts of water vapor in molecular shells and the radiation pressure on water molecules may need more detailed calculations.

    As described in the previous section, W~Hya exhibits SiO maser emission at a distance of about 0.9 times the equivalent UD diameter. The SiO maser spots are therefore co-located with the close aluminum oxide dust shell. This is very similar to what was found for S~Ori (Wittkowski et al.~\cite{Wittkowski_et07}), and supports an explanation by a similar model. 

    From the analysis of the spectrum and the fact that the flux contribution of the star/layer is below one, it was concluded that a circumstellar silicate dust shell exists around W~Hya with a lower limit to the condensation radius of about 28~times the photospheric radius. At these large radii, aluminum oxide dust grains are no longer detectable, since silicon and magnesium are much more abundant than aluminum, and they dominate the dust opacities. The low flux contribution of the silicate dust shell is consistent with a low mass-loss rate of W~Hya.

    With the previous results, the shape of the visibility curves in the left hand panel of Fig.~\ref{FigUVall} can be understood qualitatively. The visibility increase between 8 and 10~$\mu$m corresponds to the partially resolved stellar disk (including the close molecular layers) at an almost constant FDD diameter, while in the region between 10 and 12~$\mu$m, the flux contribution of the spatially resolved Al$_2$O$_3$ dust shell becomes more relevant. This leads longward of 10~$\mu$m to the nearly flat visibility curve and an increased FDD diameter (cf.~Ohnaka et al.~\cite{Ohnaka_et05} and Wittkowski et al.~\cite{Wittkowski_et07}). In addition, extinction of the outer silicate dust shell becomes important. The comparison with RR~Sco and S~Ori suggests that the partially resolved molecular layers are optically thick and that the nearby Al$_2$O$_3$ dust shell is optically thin.

    It is clear that quantitative modeling is needed to support the above findings, in particular, if the derived constituents of the close molecular and dust layers (H$_2$O, SiO?, Al$_2$O$_3$) can really provide sufficient opacities to explain the observed diameter dependence on wavelength, in particular if Al$_2$O$_3$ could cause the apparent diameter increase beyond 10~$\mu$m. To quantify the results, it will be necessary to apply dynamic atmospheric models of e.g.~Ireland \& Scholz (\cite{Ireland_Scholz06}). Even if Al$_2$O$_3$ dust has not been consistently included in these models so far, and has often to be added ad hoc\footnote{Dynamic atmospheric models, including both the molecular layers and close dust formation, are in development (Scholz and Ireland, private communication).}. This will give more detailed insight into the physical processes at work there.

%###########################################################################################
%###########################################################################################
\section{Asymmetry, intracycle variations, and cycle-to-cycle variations}\label{secPhase}

    Since observations at similar pulsation phases were mostly conducted at similar position angles, the effects of different diameters, owing to asymmetry and pulsation, are unfortunately not easy to disentangle. At visual minimum, the position angles (PA) cluster at around 90$^\circ$ (green tick marks in Fig.~\ref{FigLight_p}). At these position angles, the visibilities are higher than expected for a circular FDD, i.e.,~a smaller diameter is observed. At visual maximum, the PA is around 50$^\circ$ (red tick marks in Fig.~\ref{FigLight_p}) with the result of getting lower visibilities as expected for a circular FDD, i.e.,~a larger diameter. This trend is more clearly shown in the right hand panel of Fig.~\ref{Fig3D}, where the visibilities of the observations and the visibilities of a circular FDD are compared as a function of position angle. Obviously, the difference between both quantities changes its sign by going from 0$^\circ$ to 90$^\circ$.

    The above behavior is notable if all 75 observations over all three pulsation cycles are included in the investigation. A careful analysis, as described below, reveals both effects in W~Hya. The result will be that the diameter variation due to an elliptical asymmetry and due to the pulsation effect have the same order of magnitude.
    
    The pulsation cycle is divided into four bins as shown in Fig.~\ref{FigLight_p}. Bins~1, 2, and 3 consist of 23, 42, and 10 observations with phase ranges $0.875-0.125$, $0.125-0.375$, and $0.375-0.625$, respectively. There are no observations for bin~4 with a phase range of $0.625-0.875$. Bin~1 contains observations at visual maximum and bin~3 observations at visual minimum.
    
%%%%%%%%%%%%%%%%%%%%%%%%%%%%%%%%%%%%%%%%%%%%%%%%%%%%%%%%%%%%%%%%%%%%%%%%%%%%%%%%%%%%%%%%%%%%
\subsection{Elliptical asymmetry}\label{secPhaseSubAsy}

%---------------------------------------------------------------
   \begin{figure*}
     \centering
     \includegraphics[bb=100 0 1020 750,width=0.49\linewidth]{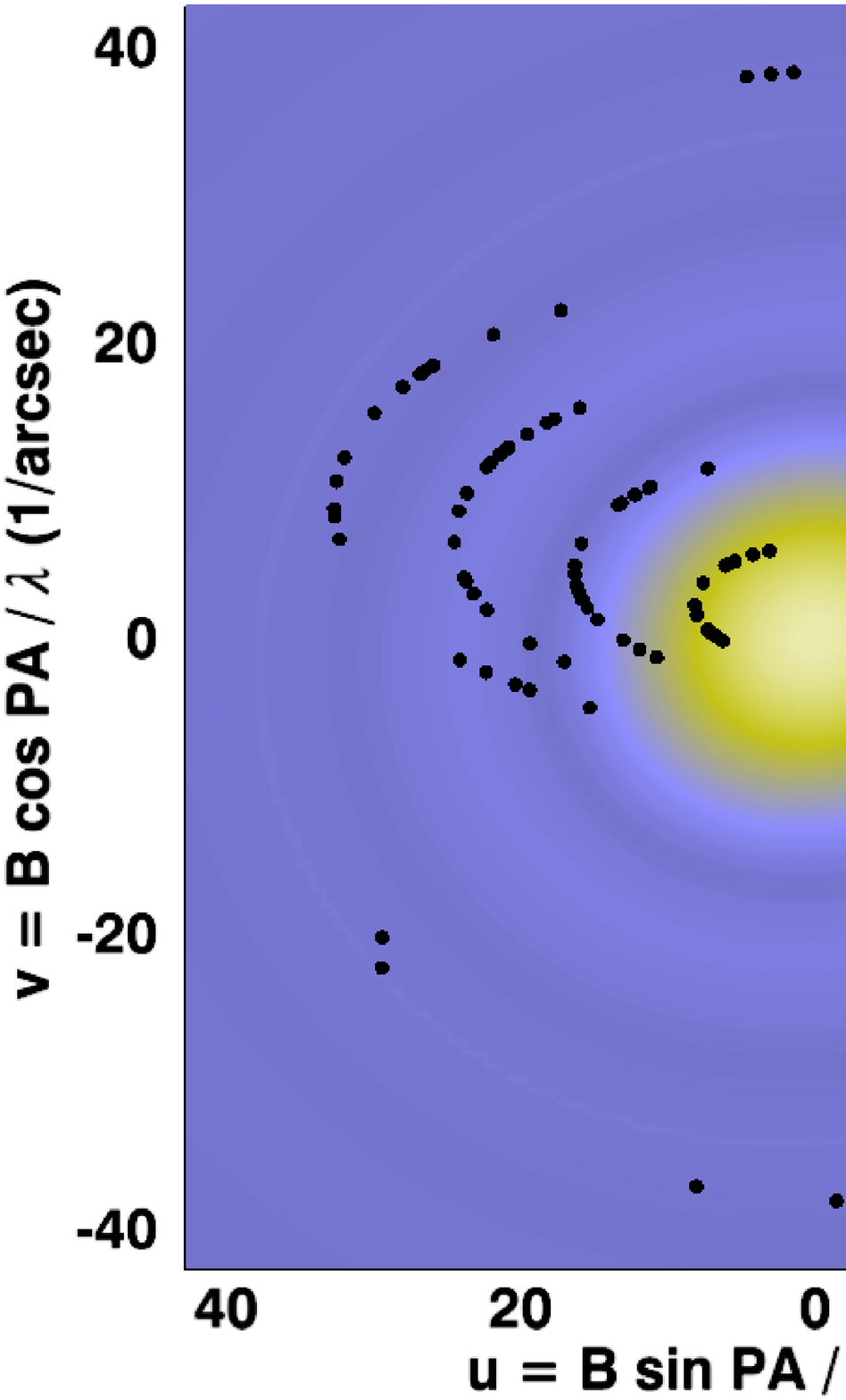}
     \includegraphics[bb=100 0 1020 750,width=0.49\linewidth]{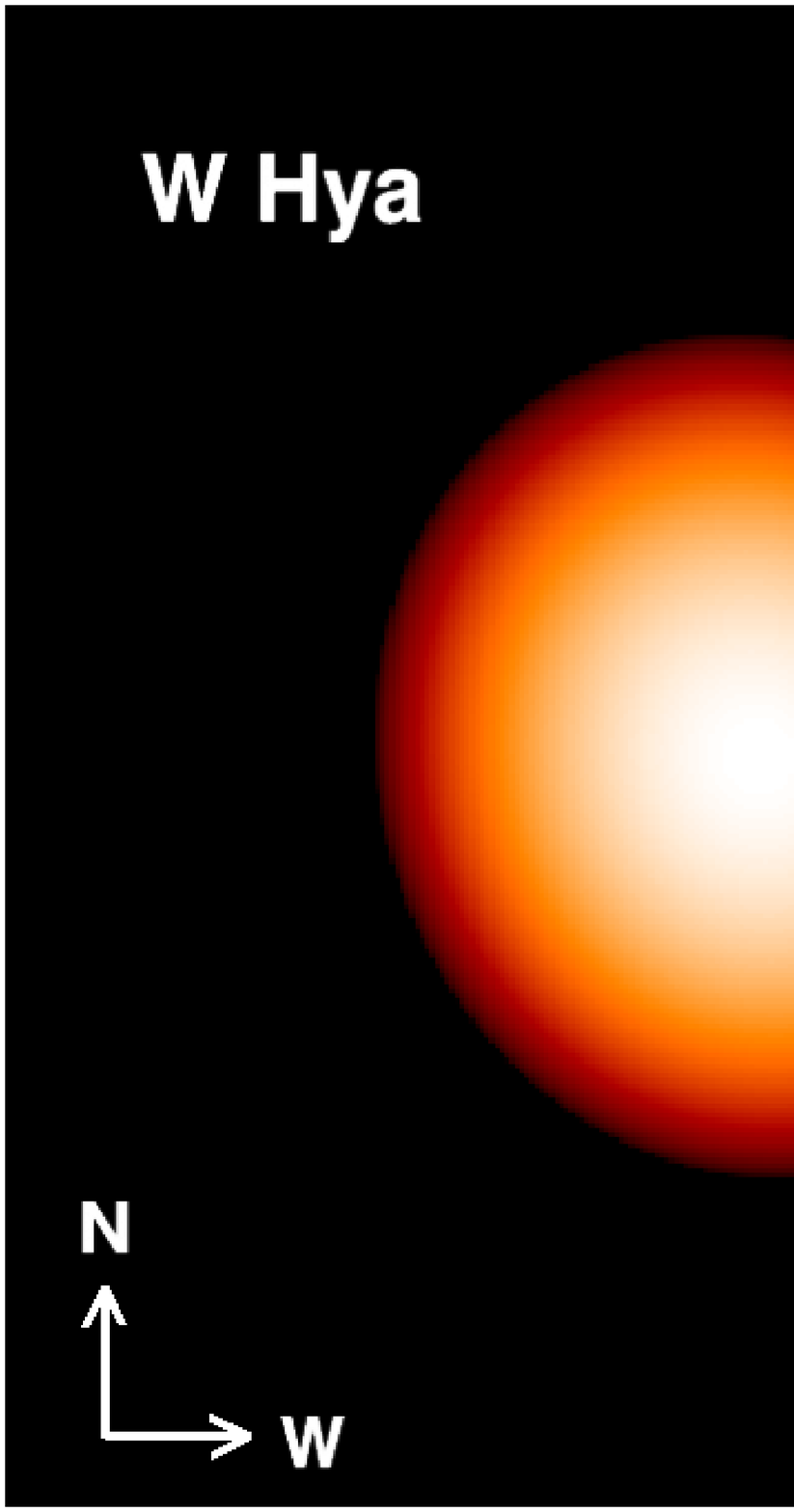}
     \caption{Asymmetric model of W~Hya. Shown are the elliptical FDD fits for wavelength bin~6 (9.07~$\mu$m). \textit{Left:} The model visibility (yellow = high visibility, blue = low visibility) plotted together with the data points in the uv-plane. The major axis in Fourier space has a position angle of 105$^\circ$. \textit{Right:} The corresponding image with a position angle of the major axis of 15$^\circ$ on the sky (linear intensity scale in arbitrary units). North is up and east is left. The size is 150~mas~$\times$~150~mas.}
     \label{FigAsymModel}
   \end{figure*}
%---------------------------------------------------------------

    Studying asymmetric features requires a good angular uv-coverage at each projected baseline length obtained within a reasonable time. Since the dependence on the visual light phase introduces already additional complexity into the model, only a simple elliptical model can be applied in order to get an indication for a departure from symmetry.    
    
    For these reasons, an elliptical FDD is fitted to subsets of the full data set. Each subset contains data with similar spatial frequencies and a narrow visual light phase range at a certain pulsation cycle. For all these configurations the resulting fits show similar nonuniformity with comparable position angles and axis ratios. The fit was repeated for the full data set to obtain an overall departure from symmetry and to have a more reliable uncertainty estimation, since the subsets have a high scatter and can contain discordant values due to fewer data points. The similarity of the results for the subsets shows that the ellipticity is stable to first order over a few years, over a certain stellar extension, and not dependent on the pulsation phase. Therefore, the parameters obtained for the full data set are given in Table~\ref{TableResults}.

    Listed are the major diameters, $\theta_{\mathrm{FDD,maj}}$, the axis ratios, $\eta$, the position angles, $\vartheta$, and the relative fluxes, $\epsilon$. The mean $\vartheta$ and $\eta$ over the full wavelength range are (11.2~$\pm$~6.2)$^\circ$ and (0.87~$\pm$~0.07), respectively. Compared to the symmetric FDD, the fluxes are identical within the errors as expected, while the reduced chi square estimates gave lower values, thus indicating a slightly better model. The right hand panel of Fig.~\ref{FigSymModel} shows the fit of the elliptical FDD for wavelength bin~6 (9.07~$\mu$m). The visibility model curves for the minor and major axes are drawn together with the data points, showing that they fit the measurements at the corresponding angles.

    However, the parameter errors are most probably underestimated because of the high number of free parameters in combination with not negligible uncertainties in the data. In particular, the PA estimation is not very well constrained, and the error might be more on the order of 15$^\circ$ to 20$^\circ$, if compared with all fitted subsets. Figure~\ref{FigAsymModel} shows the asymmetric appearance of W~Hya. The left hand panel shows the fit of the elliptical FDD for a representative wavelength, and the right hand panel displays the corresponding appearance on the sky for this model and mid-IR wavelength. The investigation of the differential phases obtained with MIDI are unfortunately not conclusive, since they are smaller than the assumed errors, and no additional constrains can be inferred from them.

\begin{figure*}
     \centering
     \includegraphics[width=0.9\linewidth]{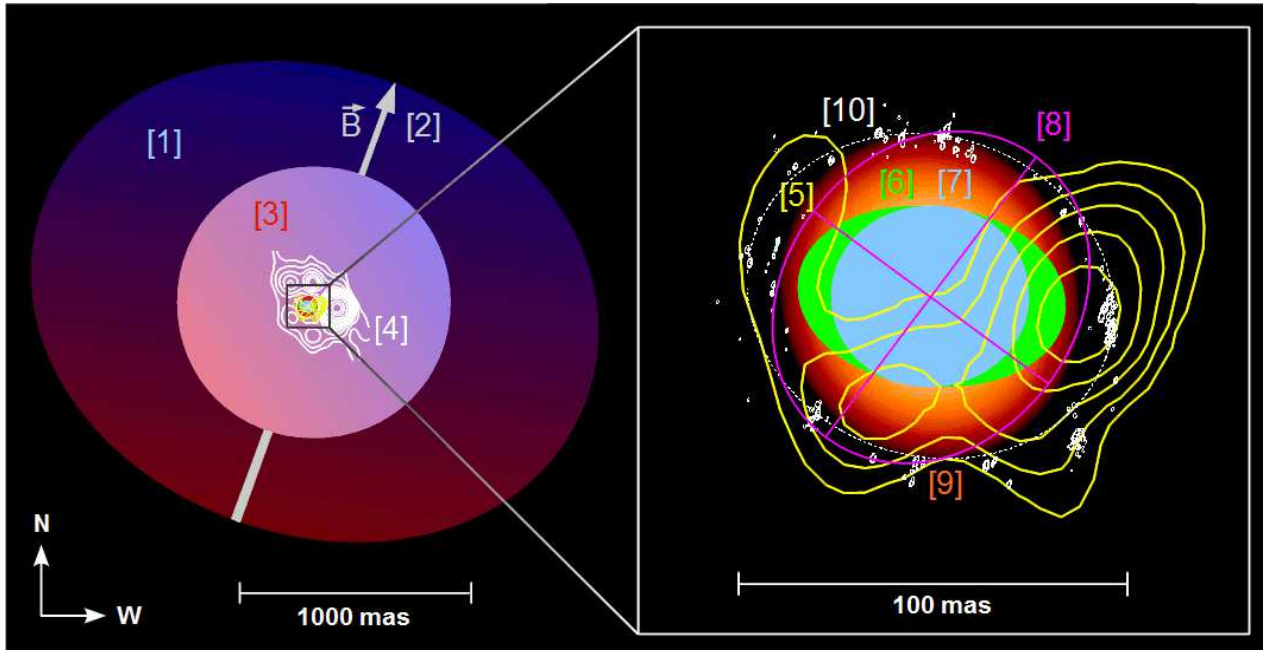}
     \caption{Asymmetry and maser measurements of W~Hya reported by various authors. See Table~\ref{TablePA} for references and text for explanation.}
     \label{FigGlobal}
   \end{figure*}

    Position angle estimations in the literature are very contradictory and are summarized in Table~\ref{TablePA} and illustrated in Fig.~\ref{FigGlobal}. Lattanzi et al.~(\cite{Lattanzi97})\footnote{Since observations were obtained only on two perpendicular baselines, the true major axis cannot be recovered.} and Tuthill et al.~(\cite{Tuthill99}) have found in the visual a PA of 143$^\circ$ and (94~$\pm$~33)$^\circ$, respectively, while no asymmetries could be detected in the near IR within the measurement uncertainties (Ireland et al.~\cite{Ireland04} and Monnier et al.~\cite{Monnier04}). From the photospheric radio continuum, Reid~\& Menten~(\cite{Reid_Menten07}) determine a PA of (83~$\pm$~18)$^\circ$, whereas Szymczak et al.~(\cite{Szymczak_et98}) find for the projected angle of the magnetic field on the plane of the sky, obtained from OH maser observations, a value of about $-20^\circ$ (the PA of the underlying OH maser distribution is 70$^\circ$), and a velocity discontinuity along the N$\rightarrow$S axis. The 215.2~GHz SO line, observed by Vlemmings et al.~(\cite{Vlemmings_et11}), also shows an velocity gradient along an N$\rightarrow$S axis with an PA of the structure of (3~$\pm$~10)$^\circ$. Muller et al.~(\cite{Muller_et08}) detect in the HCN velocity an NW$\rightarrow$SE gradient. Interestingly, the outer dust envelope, observed by Marengo et al.~(\cite{Marengo_et00}) at 18~$\mu$m, shows a very similar N$-$S elongation as determined in this work. Asymmetries in the SiO and H$_2$O maser spot distributions were, e.g., reported by Reid \& Menten (\cite{Reid_Menten90},\cite{Reid_Menten07}) and Imai et al.~(\cite{Imai_et10}).

%---------------------------------------------------------------
%---------------------------------------------------------------
%---------------------------------------------------------------
   \begin{table*}
     \caption{Position angle (PA) determinations by various authors (see Fig.~\ref{FigGlobal} for an illustration).}
     \label{TablePA}
     \centering
     \begin{tabular}{lclccc}
       \hline
       \noalign{\smallskip}
       Author & Ref$^{\mathrm{a}}$ & Obs Date & PA ($^\circ$) & Wavelength & Region$^{\mathrm{b}}$ \\
       \noalign{\smallskip}
       \hline
       \noalign{\smallskip}
         Lattanzi et al.~(\cite{Lattanzi97})   & [8] & 1995 Dec 17   & 143$^{\mathrm{c}}$   & 583 nm                   & inner \\
         Tuthill et al.~(\cite{Tuthill99})     & [7] & 1993 Jun      & 94~$\pm$~33          & 700 and 710~nm           & inner \\
         Ireland et al.~(\cite{Ireland04})     & $-$ & 2001 Feb 8/9  & $-^{\mathrm{d}}$     & $680-940$~nm             & inner \\
         Monnier et al.~(\cite{Monnier04})     & $-$ & 2000 Jan/Feb  & $-^{\mathrm{d}}$     & 2.25~$\mu$m (K band)     & inner \\
         this work                             & [9] & $2007-2009$   & 11~$\pm$~20          & $8-12$~$\mu$m (N band)   & inner \\
         Reid \& Menten (\cite{Reid_Menten07}) & [6] & 2000 Oct/Nov  & 83~$\pm$~18          & 43~GHz (radio continuum) & inner \\
                                               & [5] & 2000 Oct/Nov  & contours             & 43~GHz (SiO maser)       & inner \\
         Imai et al.~(\cite{Imai_et10})        & [10] & 2009 Feb     & spots                & 43~GHz (SiO maser)       & inner \\
         Reid \& Menten (\cite{Reid_Menten90}) & [4] & 1990 Feb      & contours             & 22~GHz (H$_2$O maser)    & intermediate \\
         Marengo et al.~(\cite{Marengo_et00})  & $-^{\mathrm{e}}$ & 1999 Jun      & N$-$S & $18$~$\mu$m (1.18''$\times$1.45'' $FWHM$)  & outer \\
         Szymczak et al.~(\cite{Szymczak_et98})& [1] & 1996 Jan 15   & N$\rightarrow$S      & 1667~MHz (OH maser), velocity gradient   & outer \\
                                               & [2] & 1996 Jan 15   & $-20$~$\pm$~20$^{\mathrm{f}}$ & 1667~MHz (OH maser), magnetic field & outer \\
         Muller et al.~(\cite{Muller_et08})    & [3] & 2008 Apr 13/15& NW$\rightarrow$SE    & 266~GHz (thermal HCN), velocity gradient & outer \\
         Vlemmings et al.~(\cite{Vlemmings_et11})& $-^{\mathrm{e}}$ & 2008 Jul 20 & 3~$\pm$~10     & 215~GHz (SO $5_5-4_4$)          & outer \\
       \noalign{\smallskip}
       \hline
     \end{tabular}
    \begin{flushleft}
       $^{\mathrm{a}}$~numbering used in Fig.~\ref{FigGlobal}; 
       $^{\mathrm{b}}$~probed regions: inner~=~$50-100$~mas, outer~$\approx$~1000~mas; 
       $^{\mathrm{c}}$~the true major axis cannot be recovered since observations are at two perpendicular baselines; 
       $^{\mathrm{d}}$~no departures from symmetry could be detected within the measurement uncertainties; 
       $^{\mathrm{e}}$~not shown in Fig.~\ref{FigGlobal} since the whole structure does not fit into the illustration; 
       $^{\mathrm{f}}$~the position angle of the major axis of the underlying OH maser distribution is 70$^\circ$
    \end{flushleft}
   \end{table*}
%---------------------------------------------------------------

    From Fig.~\ref{FigGlobal} it can be inferred that the reported velocity gradients and the magnetic field alignment are approximately in a direction perpendicular to an E-W axis where most of the maser spots are located. If the velocity gradients originate from a general rotation of the circumstellar environment and the maser spots trace the highest molecule concentrations, then this might suggest that molecular gas is primarily ejected along this E-W polar axis. This would be also consistent with the PA of the photospheric radio continuum. The dust emission traced by the observations at $18$~$\mu$m and the N-band are approximately elongated in a more perpendicular direction to this. This might suggest that the dust is more concentrated in an equatorial disk (or ring) located along an N-S axis. Therefore, one could conclude that the molecular gas and the dust are spatially separated, i.e., they have preferred ejection directions, and the different observational wavelengths probe this.

    However, this is still very uncertain since not all PA determinations fit into this picture or deviate quite a lot from this, e.g., the PAs at optical wavelengths and the 215~GHz PA. Therefore, a strict connection between the outer structures and the close stellar regions also cannot really be established. One has to keep in mind that the concentration of maser spots along the E-W polar axis may only reflect that no velocity gradients exists along this axis and masers are traceable because of the inexistent Doppler shifts.

    Other explanations are therefore still possible, in particular if time-dependent effects are important. The wide range of measured PAs at close photospheric radii might be related to a nonradial pulsation or due to clumpy dust formation. A temporally varying non-symmetric brightness distribution could also be caused by stellar spots or large convection zones. A deformation as a result of a close companion can most probably be excluded, since no evidence for such a scenario has been observed so far. However, from the data collected here, no significant temporal trends could be derived. In particular, the MIDI observations did not show any time variations of the asymmetry over about three years. This calls for more high-resolution observations and more detailed modeling. A connection to the highly elongated structures seen in planetary nebulae cannot be drawn, since W~Hya is believed to be still in an early AGB phase.

%%%%%%%%%%%%%%%%%%%%%%%%%%%%%%%%%%%%%%%%%%%%%%%%%%%%%%%%%%%%%%%%%%%%%%%%%%%%%%%%%%%%%%%%%%%%
\subsection{Intracycle variations}\label{secPhaseSubLight}
  
    To estimate a light phase dependent angular diameter in the presence of a position angle dependence, the projected baseline values of the input data are transformed in a way such that the elliptical model transforms into a circular model. This means that the projected baselines are artificially shortened (around the major axis) or lengthened (around the minor axis), i.e.~as function of position angle, so that an elliptical model fit to the data shows no departure from circular symmetry in the new coordinate system. The parameters derived from the elliptical FFD fit are used for this shearing. This transformation keeps the fitted absolute diameter values from being meaningful, so only differences with respect to a fit of a circular FDD to the full sheared data set are given.

    The fit of a circular FDD to the sheared data of the intermediate phase bin (bin~2) gives on average a (1.2~$\pm$~2.3)~mas smaller diameter. The difference between maximum phase, with an on average (2.2~$\pm$~2.3)~mas larger diameter, and minimum phase, with an on average (3.2~$\pm$~3.7)~mas smaller diameter, is (5.4~$\pm$~1.8)~mas (cf.~left hand panel of Table~\ref{TableVar}). This corresponds to a percentage change of (6~$\pm$~2)\%. The individual wavelength-dependent trends can be seen in the left hand panel of Figure~\ref{FigPara}. As can be inferred from the plot, the shapes for all individual phase bins are very similar, indicating again that the different constituents probed at different wavelengths behave similarly. The relative flux was fixed to the value obtained for the full data set (shown in the right hand panel of Fig.~\ref{FigPara}).

%---------------------------------------------------------------
   \begin{table*}
     \caption{Intracycle (\textit{left}) and cycle-to-cycle (\textit{right}) variations.}
     \label{TableVar}
     \centering
     \begin{tabular}{ccccc}
       \hline
       \noalign{\smallskip}
       Phase                & Phase & No of & Absolute$^{\mathrm{c}}$ FDD & Relative$^{\mathrm{c}}$ FDD \\
       ~Bin$^{\mathrm{a}}$  & Range &  Obs. &        Difference (mas)     &   Diameter                  \\
       \noalign{\smallskip}
       \hline
       \noalign{\smallskip}
         1   &  0.875$-$0.125 & 23 & $+$(2.2~$\pm$~2.3) & 1.02~$\pm$~0.02 \\
         2   &  0.125$-$0.375 & 42 & $-$(1.2~$\pm$~2.3) & 0.99~$\pm$~0.02 \\
         3   &  0.375$-$0.625 & 10 & $-$(3.2~$\pm$~3.7) & 0.96~$\pm$~0.04 \\
         4   &  0.625$-$0.875 &  0 &      --            &     --          \\
       \noalign{\smallskip}
       \hline
     \end{tabular}
     \hspace{1cm}
     \begin{tabular}{cccc}
       \hline
       \noalign{\smallskip}
       Cycle$^{\mathrm{b}}$ & No of & Absolute$^{\mathrm{c}}$ FDD & Relative$^{\mathrm{c}}$ FDD \\
                            &  Obs. &        Difference (mas)     &   Diameter                  \\
       \noalign{\smallskip}
       \hline
       \noalign{\smallskip}
         ~~1   &  8 & $-$(2.9~$\pm$~2.4) & 0.97~$\pm$~0.02 \\
         ~~2   &  8 & $-$(3.1~$\pm$~2.4) & 0.96~$\pm$~0.02 \\
         ~~3   & 11 & $+$(1.5~$\pm$~3.0) & 1.02~$\pm$~0.03 \\
       \noalign{\smallskip}
       \hline
       \noalign{\smallskip}
       \noalign{\smallskip}
       \noalign{\smallskip}
     \end{tabular}
     \hspace{2cm}
    \begin{flushleft}
       $^{\mathrm{a}}$~cf.~Fig.~\ref{FigLight_p}; 
       $^{\mathrm{b}}$~cf.~Fig.~\ref{FigLight_t}; 
       $^{\mathrm{c}}$~in respect to the full data set value
    \end{flushleft}
    % \begin{list}{}{}
    % \item[$^{\mathrm{a}}$] cf.~Fig.~\ref{FigLight_p};~~$^{\mathrm{b}}$~~cf.~Fig.~\ref{FigLight_t};~~$^{\mathrm{c}}$~~in respect to the full data set value
    % \end{list}
   \end{table*}
%---------------------------------------------------------------

    The diameter variation is relatively small compared to the findings by Wittkowski et al.~(\cite{Wittkowski_et07}) for the Mira variable S~Ori. A phase-dependent diameter of S~Ori cannot be derived directly from their measurements, since observations at visual maximum and visual minimum were executed at different projected baselines. However, from the modeling results shown in their Fig.~12, it can be roughly estimated that the photospheric diameter changes by approximately (15~$\pm$~5)\% and show a similar behavior as function of visual phase. The smaller variation for W~Hya can be expected since it has a smaller visual amplitude and is classified as an SRa/Mira variable, but it could also be related to the large phase-binning used in this study.

    The observed smaller angular diameter at visual minimum and the larger angular diameter at visual maximum can be explained by the phase-dependent presence of water vapor (Matsuura et al.~\cite{Matsuura_et02}) and aluminum oxide dust (Ireland \& Scholz \cite{Ireland_Scholz06} and Wittkowski et al.~\cite{Wittkowski_et07}). The star is hotter at visual maximum and H$_2$O and Al$_2$O$_3$ dust can only exist farther out. At visual minimum, dust forms closer to the stellar surface and in larger amounts (otherwise it would be not detectable with MIDI close to the star), and suggests that the mass-loss rate is higher in this phase or shortly after this phase. 

    While the observations made in this work show that possible Al$_2$O$_3$ dust exists closer to the star at visual minimum and farther out at visual maximum, it cannot be determined conclusively how this relates to the acceleration of the wind and the low mass-loss rate. Al$_2$O$_3$ can exist in the upper atmosphere without mass loss, and potentially relevant micrometer-sized Fe-free silicates were not detected with MIDI due to the low mass-loss rate and a therefore low abundance. Therefore, it is still possible that the wind formation and mass-loss mechanism is to a certain amount decoupled from the pulsation, even if it can be speculated that the relevant constituents for the wind acceleration are formed along with Al$_2$O$_3$.

    Simultaneous observations, tracing the relevant constituents and providing kinematic information, are necessary (e.g., high-resolution radio and mm observations of molecules/masers), and a comparison with dynamic atmospheric models can only be of limited success as long as the dust formation is not included.

%%%%%%%%%%%%%%%%%%%%%%%%%%%%%%%%%%%%%%%%%%%%%%%%%%%%%%%%%%%%%%%%%%%%%%%%%%%%%%%%%%%%%%%%%%%%
\subsection{Cycle-to-cycle variations}\label{secPhaseSubCycle}

    W~Hya was observed over three consecutive pulsation cycles. For each cycle, data between phases 0.2 and 0.4 are used to search for cycle-to-cycle variations. These three ranges are shaded in Fig.~\ref{FigLight_t}. The ranges contain 8, 8, and 11~observations, respectively, and have a sufficient spatial frequency coverage. A circular FDD is fitted to these observations (which are sheared as described in Sect.~\ref{secPhaseSubLight}), while the relative flux is fixed to the value obtained for the full data set. This compensates for fewer measurements and avoids the problem of being too dependent on fewer points at low spatial frequencies (i.e.~$<$~10~arcsec$^{-1}$).

    Cycles~1 and 2 give similar diameters, with values on average of about (3.0~$\pm$~2.4)~mas below the diameter obtained for the full data set (cf.~right hand panel of Table~\ref{TableVar}). The diameter for cycle~3 is marginally higher, being on average (1.5~$\pm$~3.0)~mas above the full data set value. This means that the maximal variation due to non-repeatability of the pulsation is on the order of (5~$\pm$~4)\%, and is thus lower than intracycle variations. However, regarding the uncertainties, this is only marginally significant. The wavelength-dependent shape of the diameter does not change notably between cycles.

    This gives an order of magnitude estimate of how large pulsation instabilities in SR variables are, but also reflects the chaotic atmospheric behavior. These variations are smaller than might be expected. From models by e.g.~H{\"o}fner \& Dorfi~(\cite{HoefnerDorfi_97}), Hofmann et al.~\cite{Hofmann_et98}, Ireland et al.~(\cite{Ireland_et04c}), Ireland \& Scholz~(\cite{Ireland_Scholz06}), and Nowotny et al.~(\cite{Nowotny_et10}) for Mira variables, it can be derived that the location of mass shells and dust condensation radii can vary by up to 20\% with a characteristic mean of 10\%. However, it should be kept in mind that the movement of mass shells and dust condensation radii cannot be directly compared to changes in the atmospheric radius-density structure that is traced by interferometric observations. On the other hand, these marginal size changes are not surprising if compared with the relatively small intracycle variations and the fact that W~Hya is an SR variable. Finally, this verifies that the folding of consecutive pulsation cycles in the previous diameter and intracycle analysis is an acceptable assumption.

%###########################################################################################
%###########################################################################################
\section{Summary}\label{secConc}

    W~Hya was monitored over about three years in the thermal IR ($8-12$~$\mu$m). These are the first high-resolution interferometric N-band observations of W~Hya with MIDI. A photometric study reveals a clear phase dependency of the N-band flux with a flux variation on the order of 20\% between maximum and minimum light. The mid-IR maximum occurs after the visual maximum at visual phase 0.15~$\pm$~0.05.
    
    The visibility data can be best fitted with a fully limb-darkened disk, which accounts to some extent for surrounding atmospheric layers. The resulting FDD diameter of W~Hya is almost constant between 8 and 10~$\mu$m at a value of about (80~$\pm$~1.2)~mas (7.8~AU), while it gradually increases at wavelengths longer than 10~$\mu$m to reach (105~$\pm$~1.2)~mas (10.3~AU) at 12~$\mu$m. In contrast, the relative flux decreases from (0.85~$\pm$~0.02) to (0.77~$\pm$~0.02), reflecting the increased flux contribution from a fully resolved surrounding silicate dust shell. From field-of-view effects, it could be derived that the silicate dust shell has an inner radius larger than 28 photospheric radii ($>$50~AU, $>$0.5~arcsec).

    The measured apparent mid-IR diameter at 10~$\mu$m is about 1.6~$\pm$~0.2 times larger than the near K-band diameter ($\theta_{\mathrm{UD},10 \mu\mathrm{m}}$/$\theta_{\mathrm{UD},2.2 \mu\mathrm{m}}$), i.e.~about 1.9 times the photospheric diameter. This is very similar to findings by Weiner et al.~(\cite{Weiner_et03}), Perrin et al.~(\cite{Perrin_et04}) and Tatebe et al.~(\cite{Tatebe_et06}) for similar stars. In particular, the diameter behavior throughout the N-band is comparable with observations of the O-rich Mira stars RR~Sco and S~Ori by Ohnaka et al.~(\cite{Ohnaka_et05}) and Wittkowski et al.~(\cite{Wittkowski_et07}), respectively.
    
    W~Hya is therefore described by an analogous model. The constant diameter part results from a partially resolved stellar disk, including the close molecular layer of H$_2$O, while the increase beyond 10~$\mu$m can be most likely attributed to the contribution of a spatially resolved nearby amorphous Al$_2$O$_3$ dust shell. Probably owing to the low mass-loss rate, close Fe-free silicate dust, as proposed by H{\"o}fner~\cite{Hoefner08}, could not be detected with MIDI.

    Since observations at similar pulsation phases were mostly conducted at similar position angles, the effects of different diameters, owing to elliptical asymmetry and pulsation, are unfortunately not easy to disentangle. By only using certain pulsation phases and projected baselines an asymmetric character of the extended structure could be confirmed. An elliptical FDD, with a position angle of (11~$\pm$~20)$^\circ$ and an axis ratio of (0.87~$\pm$~0.07), fits the data. The asymmetry might be explained by an enhanced dust concentration along an N-S axis.
    
    To estimate the dependence of the angular diameter as a function of visual light phase, the input data are sheared. The observed angular diameter is smaller at visual minimum and larger at visual maximum with a periodic change of (5.4~$\pm$~1.8)~mas between maximum and minimum, corresponding to about (6~$\pm$~2)\%. This is much less than reported for the high amplitude pulsating Mira star S~Ori (Wittkowski et al.~\cite{Wittkowski_et07}). The smaller observed angular diameter at visual minimum can be explained by the phase-dependent presence of water vapor and likely aluminum oxide dust and their temperature sensitivity. Since this variation only traces the location of constituents that are probably not relevant for the wind formation, no firm conclusions can be drawn concerning the mass-loss mechanism. One can only speculate that more dust forms at visual minimum and that the mechanism for a moderate-to-low mass-loss rate is similar for O-rich SR and Mira variables. The detected cycle-to-cycle variations are smaller than intracycle variations and on the order of (5~$\pm$~4)\%.
    
    The observation of nearby molecular layers and a nearby dust shell, like in other AGB stars, confirms the emerging standard picture and supports the need for self-consistent dynamic atmospheric models with consistently included dust formation close to the star. Primarily, the close Al$_2$O$_3$ dust shell most likely detected in our observations, well below the distance at which the silicate dust shell is traced, has now been revealed in a few objects. It has also been shown that a good phase and uv-coverage over the whole pulsation cycle can be crucial, while interferometric observations in the N-band are an irreplaceable tool for resolving close stellar structures and for searching for atmospheric constituents. Future work will concentrate on improving the data reduction process further and on applying dynamical atmospheric models to the data.

%###########################################################################################
%###########################################################################################

\begin{acknowledgements}
    We thank Michael Scholz and Mike Ireland for providing new theoretical models, Markus Wittkowski and Keiichi Ohnaka for fruitful discussion, and Henry Woodroff and Mike Ireland for making diameter data sets available. The first author would also like to thank the \emph{International Max-Planck Research School} (IMPRS) for its financial support with a fellowship. We acknowledge with thanks the variable star observations from the AAVSO International Database contributed by observers worldwide and used in this research. This research has made use of the SIMBAD database, operated at the CDS, France, the ISO/IRAS database and NASA's Astrophysical Data System. We would also thank the referee for his or her very valuable comments.
\end{acknowledgements}

%###########################################################################################
%###########################################################################################

%###########################################################################################
%###########################################################################################

\end{document}